\newif\iffigs\figstrue

\documentclass[12pt]{article}
\iffigs
   \input{epsf}
\else
\fi
\newcommand{\eqn}[1]{(\ref{#1})}

\newsavebox{\uuunit}
\sbox{\uuunit}
    {\setlength{\unitlength}{0.825em}
     \begin{picture}(0.6,0.7)
        \thinlines
        \put(0,0){\line(1,0){0.5}}
        \put(0.15,0){\line(0,1){0.7}}
        \put(0.35,0){\line(0,1){0.8}}
       \multiput(0.3,0.8)(-0.04,-0.02){12}{\rule{0.5pt}{0.5pt}}
     \end {picture}}

\def\IP{\relax{\rm I\kern-.18em P}}

\begin{document}
%
\font\cmss=cmss10 \font\cmsss=cmss10 at 7pt
\def\twomat#1#2#3#4{\left(\matrix{#1 & #2 \cr #3 & #4}\right)}
\def\inbar{\vrule height1.5ex width.4pt depth0pt}
\def\IC{\relax\,\hbox{$\inbar\kern-.3em{\rm C}$}}
\def\IG{\relax\,\hbox{$\inbar\kern-.3em{\rm G}$}}
\def\IB{\relax{\rm I\kern-.18em B}}
\def\ID{\relax{\rm I\kern-.18em D}}
\def\IL{\relax{\rm I\kern-.18em L}}
\def\IF{\relax{\rm I\kern-.18em F}}
\def\IH{\relax{\rm I\kern-.18em H}}
\def\II{\relax{\rm I\kern-.17em I}}
\def\IN{\relax{\rm I\kern-.18em N}}
\def\IP{\relax{\rm I\kern-.18em P}}
\def\IQ{\relax\,\hbox{$\inbar\kern-.3em{\rm Q}$}}
\def\bfzero{\relax\,\hbox{$\inbar\kern-.3em{\rm 0}$}}
\def\IR{\relax{\rm I\kern-.18em R}}
\def\ZZ{\relax\ifmmode\mathchoice
{\hbox{\cmss Z\kern-.4em Z}}{\hbox{\cmss Z\kern-.4em Z}}
{\lower.9pt\hbox{\cmsss Z\kern-.4em Z}}
{\lower1.2pt\hbox{\cmsss Z\kern-.4em Z}}\else{\cmss Z\kern-.4em
Z}\fi}
\def\bfone{\relax{\rm 1\kern-.35em 1}}
\def\dop{{\rm d}\hskip -1pt}
\def\real{{\rm Re}\hskip 1pt}
\def\trace{{\rm Tr}\hskip 1pt}
\def\ii{{\rm i}}
\def\diag{{\rm diag}}
\def\sch#1#2{\{#1;#2\}}
\def\bfone{\relax{\rm 1\kern-.35em 1}}
\font\cmss=cmss10 \font\cmsss=cmss10 at 7pt
\def\a{\alpha} \def\b{\beta} \def\d{\delta}
\def\e{\epsilon} \def\c{\gamma}
\def\G{\Gamma} \def\l{\lambda}
\def\L{\Lambda} \def\s{\sigma}
\def\cA{{\cal A}} \def\cB{{\cal B}}
\def\cC{{\cal C}} \def\cD{{\cal D}}
\def\cF{{\cal F}} \def\cG{{\cal G}}
\def\cH{{\cal H}} \def\cI{{\cal I}}
\def\cJ{{\cal J}} \def\cK{{\cal K}}
\def\cL{{\cal L}} \def\cM{{\cal M}}
\def\cN{{\cal N}} \def\cO{{\cal O}}
\def\cP{{\cal P}} \def\cQ{{\cal Q}}
\def\cR{{\cal R}} \def\cV{{\cal V}}\def\cW{{\cal W}}
\newcommand{\be}{\begin{equation}}
\newcommand{\ee}{\end{equation}}
\newcommand{\bea}{\begin{eqnarray}}
\newcommand{\eea}{\end{eqnarray}}
\let\la=\label \let\ci=\cite \let\re=\ref
%
%
%
\def\crr{\crcr\noalign{\vskip {8.3333pt}}}
\def\tilde{\widetilde}
\def\bar{\overline}
\def\us#1{\underline{#1}}
\let\shat=\hat
\def\hat{\widehat}
\def\hyp{\vrule height 2.3pt width 2.5pt depth -1.5pt}
\def\square{\mbox{.08}{.08}}
\def\Coeff#1#2{{#1\over #2}}
\def\Coe#1.#2.{{#1\over #2}}
\def\coeff#1#2{\relax{\textstyle {#1 \over #2}}\displaystyle}
\def\coe#1.#2.{\relax{\textstyle {#1 \over #2}}\displaystyle}
\def\half{{1 \over 2}}
\def\shalf{\relax{\textstyle {1 \over 2}}\displaystyle}
\def\dag#1{#1\!\!\!/\,\,\,}
\def\to{\rightarrow}
\def\notin{\hbox{{$\in$}\kern-.51em\hbox{/}}}
\def\shdot{\!\cdot\!}
\def\ket#1{\,\big|\,#1\,\big>\,}
\def\bra#1{\,\big<\,#1\,\big|\,}
\def\equaltop#1{\mathrel{\mathop=^{#1}}}
\def\Trbel#1{\mathop{{\rm Tr}}_{#1}}
\def\inserteq#1{\noalign{\vskip-.2truecm\hbox{#1\hfil}
\vskip-.2cm}}
\def\attac#1{\Bigl\vert
{\phantom{X}\atop{{\rm\scriptstyle #1}}\phantom{X}}}
\def\exx#1{e^{{\displaystyle #1}}}
\def\del{\partial}
\def\delbar{\bar\partial}
\def\nex#1{$N\!=\!#1$}
\def\dex#1{$d\!=\!#1$}
\def\cex#1{$c\!=\!#1$}
\def\eg{{\it e.g.}} \def\ie{{\it i.e.}}
\def\IE{\relax{{\rm I\kern-.18em E}}}
\def\cE{{\cal E}}
\def\rt{{\cR^{(3)}}}
\def\IGam{\relax{{\rm I}\kern-.18em \Gamma}}
\def\IGa{\IA}
\def\ii{{\rm i}}
\begin{titlepage}
\hskip 8.5cm
\hskip 1.5cm
\vbox{\hbox{CERN-TH/96-365}\hbox{hep-th/9612202}\hbox{December, 1996}}
\vfill
\begin{center}
{\LARGE Solvable Lie Algebras \\
\vskip 1.5mm
in Type IIA, Type IIB and M Theories$^*$ }\\
\vfill
{\large Laura Andrianopoli$^1$,
Riccardo D'Auria $^2$, Sergio Ferrara$^2$,\\ Pietro Fr\'e$^3$,
Ruben Minasian$^2$ and  Mario Trigiante$^4$   } \\
\vfill
{\small
$^1$ Dipartimento di Fisica Universit\'a di Genova, via Dodecaneso 33,
I-16146 Genova\\
and Istituto Nazionale di Fisica Nucleare (INFN) - Sezione di Genova, Italy\\
\vspace{6pt}
$^2$ CERN, Theoretical Division, CH 1211 Geneva,
Switzerland,\\
\vspace{6pt}
$^3$ Dipartimento di Fisica Teorica, Universit\'a di Torino, via P. Giuria 1,
I-10125 Torino, \\
 Istituto Nazionale di Fisica Nucleare (INFN) - Sezione di Torino, Italy \\
\vspace{6pt}
$^4$ International School for Advanced Studies (ISAS), via Beirut 2-4,
I-34100 Trieste\\
and Istituto Nazionale di Fisica Nucleare (INFN) - Sezione di Trieste, Italy\\
\vspace{6pt}
}
\end{center}
\vfill
\begin{center}
{\bf Abstract}
\end{center}
{\small We study some applications of solvable Lie algebras in type $IIA$, type $IIB$ and $M$ theories.
 $RR$ and $NS$ generators  find a natural geometric interpretation in this framework.
Special emphasis is given to the counting of the abelian nilpotent ideals (translational symmetries of
the scalar manifolds) in arbitrary $D$ dimensions.
These are seen to be related, using  Dynkin diagram techniques,  to one-form counting in $D+1$ dimensions.
A recipe for gauging isometries  in this framework is also presented.
In particular, we list  the gauge groups both for compact and translational isometries. The former
agree with some results already existing in gauged supergravity.
The latter should be possibly related to the study of partial supersymmetry breaking, as suggested by a similar role
played by solvable Lie algebras in $N=2$ gauged supergravity.}
\vspace{2mm} \vfill \hrule width 3.cm
{\footnotesize
 $^*$ Supported in part by   EEC  under TMR contract
 ERBFMRX-CT96-0045, in which L. Andrianopoli and R. D'Auria are associated to Torino University
 and S. Ferrara and M. Trigiante are associated to Frascati and by DOE grant DE-FG03-91ER40662.}
\end{titlepage}
\section{Introduction}
Hidden non compact symmetries  of extended supergravities \cite{cre} have recently played a major role in unreaveling
some non perturbative properties
of string theories such as various types of dualities occurring in different dimensions and in
certain regions of the moduli spaces \cite{schw}.
\par
In particular their discrete remnants have been a crucial importance to discuss, in a model independent way, some physical properties
such as the spectrum of BPS states \cite{sesch}, \cite{hamo}, \cite{huto2} and entropy formulas for extreme black--holes
\cite{feka}\cite{malda}.
\par
It is common wisdom that such U--dualities should play an important role in the understanding
of other phenomena such as the mechanism for supersymmetry breaking, which may be due to some non perturbative physics
 \cite{postr},\cite{wi},\cite{klmvw}.
\par
Recently, we have analyzed some properties of U--duality symmetries
in any dimensions in the context of solvable Lie algebras \cite{noialtri}.
\par
In string theories or M--theory compactified to lower dimensions \cite{witten},
preserving $N>2$ supersymmetries, the U--duality group  is generically
an infinite dimensional discrete subgroup $U(\ZZ) \subset U$,
where $U$ is related to the non--compact symmetries of the low energy effective
supergravity theory \cite{huto}.
\par
The solvable Lie algebra $G_S=Solv(U/H)$ with the property $\exp [G_S] = U/H$,
where $U/H$ is (locally) the scalar manifold of the theory, associates
group generators to each scalar, so that one can speak of NS and R--R generators.
\par
Translational symmetries of NS and/or R--R fields are associated with the maximal
abelian nilpotent ideal of $\cA \subset G_S$, with a series of implications.
\par
The advantage of introducing such notion is twofold:
besides that  of associating generators with  scalar fields when decomposing
the U--duality group with respect to perturbative and non perturbative
symmetries of string theories, such as T and S--duality in type IIA
 or $SL(2,\IR)$
duality in type IIB, one may unreveal connections between different theories
and have an understanding of N--S and R--R generators at the group--theoretical
level, which may hold beyond a particular perturbative framework.
Furthermore, the identification $U/H \sim \exp[G_S]$ of the scalar coset manifold
with the group manifold of a normed solvable Lie algebra allows the description
of the local differential geometry of $U/H$ in purely algebraic terms.
Since the effective low energy supergravity lagrangian is entirely encoded in terms of this
local differential geometry, this fact has obvious distinctive advantages.
\par
In the present paper we derive a certain number of relations among
 solvable Lie algebras which explain some of the results obtained by some of
us in a previous work.
\par
In particular, we show that the Peccei--Quinn (translational) symmetries of $U_{D}/H_{D}$
in $D=10-r$ dimensions are classified  by $U_{D+1}$, while their
 NS and R--R content are classified by $O(r-1,r-1)$.
\par
For $D>3$ this content corresponds to the number of vector fields in the $D+1$
theory, at least for maximal supergravities.
\par
An explicit expression for these generators is given and an interpretation
in terms of branes is also obtained.
\par
We also compare different decompositions of solvable Lie algebras in IIA, IIB and M--theory in toroidal
compactifications which preserve maximal supersymmetry (i.e. 32 supercharges).
Solvable Lie algebras in the context of compactifications preserving lower supersymmetries will
be discussed elsewhere.
\par
While in Type IIA the relevant decomposition is with respect to the S--T
duality group, in IIB theory we decompose the U--duality group with respect
to $SL(2,\IR)\times GL(r,\IR)$ and in M--theory with respect to $GL(r+1, \IR)$.
\par
Comparison of these decompositions show some of the non-perturbative
relations existing among these theories, such as the interpretation of
$SL(2,\ZZ)$ as the group acting on the complex structure of a two-dimensional
 torus \cite{vasch}.

Solvable Lie algebras play also an important role in the gauging of isometries
while preserving vanishing cosmological constant or partially breaking
some of the supersymmetries. Indeed, this was used in the literature
 \cite{fegipo} in the
 context of $N=2$ supergravity spontaneously broken to $N=1$ and may be used
 in a more general framework. This study is relevant in view of possible applications
 in string effective field theories, where field-strength condensation
may give rise to the gauging of isometries \cite{postr}.
\par
This paper is organized as follows:
In section 2 we briefly recall the solvable Lie algebra structure
 of maximally extended supergravities in any dimensions and their
N--S and R--R content.
\par
In section 3 we study four embedding chains of subalgebras of
the maximal non compact $E_{r+1(r+1)}$  series \cite{cre} of   U--duality algebras.
 Physically these chains are  related with
the type IIA, type IIB and M--theory interpretation of maximal supergravity in
$D$--dimensions. First we focus on their algebraic characterization
using Dynkin diagram techniques and  with such analysis we
show how to represent the generators of all the relevant
solvable Lie algebras within the $E_7$ root space.
Then we consider the physical interpretation of the embedding chains
and we emphasize the perfect match between
algebraic structures and the string theory counting of massless modes.
\par
In section 4 we study the role
of the maximal abelian nilpotent ideals and their interpretation
in terms of brane wrapping and reducing.
\par
In section 5 the gauging of isometries is studied.
We show how the results of sections 3 and 4
lead to a natural filtration of the solvable Lie algebra
which provides a canonical polynomial parametrization of the
supergravity scalar coset manifold $U_D/H_D$. This parametrization
is of special value in addressing the solution of problems like
the extremization of black--hole entropy or other questions related with
supergravity central charges, besides the problem of gauging.
Then a list of maximal gaugings, both for compact and translational isometries,
 is given and is seen to agree with
some results previously obtained in gauged maximally extended supergravities.
\par
In section six we end with some concluding remarks.
 \par
The one--to--one identification of scalar fields with the generators of the
solvable algebra is given
in appendix A while the representation matrices of the gauged abelian ideals
for all dimension $4 \le D \le 9$ is given in
appendix B.

\section{The solvable Lie algebra structure: NS and RR scalar fields}
It has been known for many years \cite{sase} that the scalar field manifold of
both pure and matter coupled
$N>2$ extended supergravities in $D=10-r$ ($r=6,5,4,3,2,1$) is a non compact
homogeneus symmetric manifold $U_{(D,N)} /H_{(D,N)}$, where $U_{(D,N)}$ (depending
on the space--time dimensions and on the number of supersymmetries) is a non compact
Lie group and $H_{(D,N)}\subset U_{(D,N)}$ is a maximal compact subgroup.
Furthermore, the structure of the supergravity lagrangian is completely encoded in the
local differential geometry of
$U_{(D,N)}/H_{(D,N)}$, while an appropriate restriction to integers $U_{(D,N)}(\ZZ)$ of
the Lie group $U_{(D,N)}$
is the conjectured U--duality symmetry of string theory that unifies  T--duality with S--duality \cite{huto}.
\par
As we discussed in a recent paper \cite{noialtri}, utilizing a well established mathematical
framework \cite{helgason}, in all these cases the scalar coset manifold $U/H$ can be
identified with the group manifold of a normed solvable Lie algebra:
\begin{equation}
  U/H \sim \exp[{Solv}]
\end{equation}

The representation of the supergravity scalar manifold ${\cal M}_{scalar}= U/H$
as the group manifold associated with a {\it  normed solvable Lie algebra}
introduces a one--to--one correspondence between the scalar fields $\phi^I$ of
supergravity and the generators $T_I$ of the solvable Lie algebra $Solv\, (U/H)$.
Indeed the coset representative $L(U/H)$ of the homogeneous space $U/H$ is
identified with:
\begin{equation}
L(\phi) \, =\, \exp [ \phi^I \, T_I ]
\label{cosrep1}
\end{equation}
where $\{ T_I \}$ is a basis of $Solv\, (U/H)$.
\par
As a consequence of this fact the tangent bundle to the scalar manifold $T{\cal M}_{scalar}$
is identified with the solvable Lie algebra:
\begin{equation}
T{\cal M}_{scalar} \, \sim \,Solv \, (U/H)
\label{cosrep2}
\end{equation}
and any algebraic property of the solvable algebra has a corresponding physical interpretation in
terms of string theory massless field modes.
\par
Furthermore, the local differential geometry of the scalar manifold is described
 in terms of the solvable Lie algebra structure.
Given the euclidean scalar product on $Solv$:
\begin{eqnarray}
  <\, , \, > &:& Solv \otimes Solv \rightarrow \IR
\label{solv1}\\
<X,Y> &=& <Y,X>\label{solv2}
\end{eqnarray}
the covariant derivative with respect to the Levi Civita connection is given by
the Nomizu operator \cite{alex}:
\begin{equation}
\forall X \in Solv : \IL_X : Solv \to Solv
\end{equation}
\begin{eqnarray}
  \forall X,Y,Z \in Solv & : &2 <Z,\IL_X Y> \nonumber\\
&=& <Z,[X,Y]> - <X,[Y,Z]> - <Y,[X,Z]>
\label{nomizu}
\end{eqnarray}
and the Riemann curvature 2--form is given by the commutator of two Nomizu
operators:
\begin{equation}
 <W,\{[\IL_X,\IL_Y]-\IL_{[X,Y]}\}Z> = R^W_{\ Z}(X,Y)
\label{nomizu2}
\end{equation}
In the case of maximally extended supergravities in $D=10-r$ dimensions the scalar
manifold has a universal structure:
\begin{equation}
 { U_D\over H_D}  = {E_{r+1(r+1)} \over H_{r+1}}
\label{maximal1}
\end{equation}
where the Lie algebra of the $U_D$--group $E_{r+1(r+1)} $ is the
maximally non compact real section of the exceptional $E_{r+1}$ series  of the simple complex
Lie Algebras
and $H_{r+1}$ is its maximally compact subalgebra \cite{cre}.
As we discussed in a recent paper \cite{noialtri},
the manifolds $E_{r+1(r+1)}/H_{r+1}$
share the distinctive  property of being non--compact homogeneous spaces of maximal rank
$r+1$, so that the associated solvable Lie algebras,
 such that ${E_{r+1(r+1)}}/{H_{r+1}} \, = \, \exp \left [ Solv_{(r+1)} \right ]
$,  have the particularly simple structure:
\begin{equation}
Solv\, \left ( E_{r+1}/H_{r+1} \right )\, = \, {\cal H}_{r+1} \, \oplus_{\alpha \in
\Phi^+(E_{r+1})} \, \IE^\alpha
\label{maxsolv1}
\end{equation}
where $\IE^\alpha \, \subset \, E_{r+1}$ is the 1--dimensional subalgebra associated
with the root $\alpha$
and $\Phi^+(E_{r+1})$ is the positive part of the $E_{r+1}$--root--system.
\par
The generators of the solvable Lie algebra  are in
one--to--one correspondence with the scalar fields of the theory.
Therefore they can be characterized as Neveu--Schwarz or Ramond--Ramond
depending on their origin in compactified string theory. From the
algebraic point of view the generators of the solvable algebra are of
three possible types:
\begin{enumerate}
\item {Cartan generators }
\item { Roots that belong to the adjoint representation of the
$D_r \equiv SO(r,r) \subset E_{r+1(r+1)}$ subalgebra (= the T--duality algebra) }
\item {Roots which are weights of an irreducible representation
 of the $D_r$ algebra.}
\end{enumerate}
The scalar fields associated with generators of type 1 and 2 in the above
list are Neveu--Schwarz fields while the fields of type 3 are
Ramond--Ramond fields.
\par
In the $r=6$ case, corresponding to $D=4$, there is one extra root,
besides those listed above, which is also of the Neveu--Schwarz type.
From the dimensional reduction viewpoint the origin of this extra
root is the following: it is associated with the axion $B_{\mu\nu}$
which only in 4--dimensions becomes equivalent to a scalar field.
This root (and its negative) together with the 7-th Cartan generator
of $O(1,1)$ promotes the S--duality in $D=4$ from $O(1,1)$, as it is in
all other dimensions, to $SL(2,\IR)$.


\subsection{Counting  of massless modes in sequential toroidal compactifications
of $D=10$ type IIA superstring}
In order to make the pairing between scalar field modes and solvable Lie algebra generators
explicit, it is convenient to organize the counting of bosonic zero modes
in a sequential way that goes down from $D=10$ to $D=4$ in 6 successive steps.

The useful feature of this sequential viewpoint is that it has a direct algebraic
counterpart in the successive embeddings of the exceptional Lie Algebras $E_{r+1}$
one into the next one:
{\footnotesize
\begin{equation}
  \matrix{E_{7(7)}&\supset &  E_{6(6)}&\supset & E_{5(5)}&\supset & E_{4(4)}&\supset
  & E_{3(3)}&\supset & E_{2(2)}&\supset & O(1,1) \cr
D=4 & \leftarrow & D=5 & \leftarrow & D=6 & \leftarrow & D=7 & \leftarrow & D=8 &
\leftarrow & D=9 & \leftarrow & D=10 \cr}
\end{equation}}
If we consider the bosonic massless spectrum \cite{gsw} of  type II theory in $D=10$
in the Neveu--Schwarz sector we have the metric, the axion and the dilaton,
while in the Ramond--Ramond sector we have a 1--form and a 3--form:
\begin{equation}
 D=10 \quad : \quad  \cases{
 NS: \quad g_{\mu\nu}, B_{\mu\nu} , \Phi \cr
 RR: \quad  A_{\mu} , A_{\mu\nu\rho} \cr}
 \label{d10spec}
\end{equation}
corresponding to the following counting of degrees of freedom:
$\# $ d.o.f. $g_{\mu\nu} = 35$, $\# $ d.o.f. $B_{\mu\nu} = 28$, $\# $ d.o.f. $A_{\mu} = 8$, $\# $ d.o.f. $A_{\mu\nu\rho} = 56$
so that the total number of degrees of freedom is $64$  both in the Neveu--Schwarz
and in the Ramond:
 \begin{eqnarray}
  \mbox{Total $\#$ of NS degrees of freedom}&=&{ 64}={ 35}+{ 28}+ { 1} \nonumber\\
  \mbox{Total $\#$ of RR degrees of freedom}&=&{ 64}={ 8}+{ 56}
  \label{64NSRR}
 \end{eqnarray}
 \par
It is worth noticing that the number of degrees of freedom of N--S and R--R sectors are equal, both for
bosons and fermions, to $128= (64)_{NS} + (64)_{RR}$. This is merely a consequence
of type II supersymmetry.
Indeed, the entire Ramond sector (both in type IIA and type IIB) can be thought as a spin $3/2$
multiplet of the second supersymmetry generator.
\par
Let us now organize the degrees of freedom as they appear after toroidal compactification
on a $r$--torus \cite{pope}:
\begin{equation}
{\cal M}_{10} = {\cal M}_{D-r} \, \otimes T_r
\end{equation}
Naming with Greek letters the world indices on the $D$--dimensional
space--time and with Latin letters the internal indices referring to
the torus dimensions we obtain the results displayed in Table \ref{tabu1} and number--wise we
obtain the counting of Table \ref{tabu2}:
\par
\vskip 0.3cm
\begin{table}[ht]
\begin{center}
\caption{Dimensional reduction of type IIA fields}
\label{tabu1}
 \begin{tabular}{|l|l|c|c|c|c|r|}\hline
\vline & \null & \vline & Neveu Schwarz & \vline & Ramond Ramond & \vline \\
 \hline
 \hline
\vline & Metric & \vline &  $g_{\mu\nu}$& \vline  & \null & \vline   \\ \hline
\vline & 3--forms & \vline &  \null & \vline  & $A_{\mu\nu \rho}$ & \vline \\ \hline
\vline & 2--forms & \vline   & $B_{\mu\nu}$ & \vline   & $A_{\mu\nu i}$ & \vline  \\ \hline
\vline & 1--forms & \vline   &  $g_{\mu i}, \quad B_{\mu i}$ & \vline  & $A_{\mu},
 \quad A_{\mu ij}$ & \vline \\ \hline
\vline & scalars  & \vline  & $\Phi, \quad g_{ij}, \quad B_{ij}$ &
 \vline  & $A_{i}, \quad A_{ijk}$ & \vline \\ \hline
 \end{tabular}
 \end{center}
\end{table}
 \par
 \vskip 0.3cm
 \par
\vskip 0.3cm
\begin{table}[ht]
\begin{center}
\caption{Counting of type IIA fields}

\label{tabu2}
 \begin{tabular}{|l|l|c|c|c|c|r|}\hline
\vline & \null & \vline & Neveu Schwarz & \vline & Ramond Ramond & \vline \\
 \hline
 \hline
\vline & Metric & \vline &  ${ 1}$& \vline  & \null & \vline   \\ \hline
\vline & $\#$ of 3--forms & \vline &  \null & \vline  & ${  1}$ & \vline \\ \hline
\vline &$\#$ of  2--forms & \vline   & ${  1}$ & \vline   & ${  r}$ & \vline  \\ \hline
\vline &$\#$ of 1--forms & \vline   &  $ {  2 r}$ & \vline  & $ {  1} +
\frac 1 2 \, r \, (r-1)$ & \vline \\ \hline
\vline & scalars  & \vline  & $1 \, +  \, \frac 1 2 \, r \, (r+1)$&
\vline  & $ r \, +\, \frac 1 6 \, r \, (r-1) \, (r-2)  $ & \vline \\
\vline &  \null   & \vline & $  +  \, \frac 1 2 \, r \, (r-1)  $ & \vline & \null   & \vline \\
\hline
\end{tabular}
\end{center}
\end{table}
\vskip 0.2cm
We can easily check that the total number
of degrees of freedom in both sectors is indeed $64$ after
dimensional reduction as it was before.

\section{ $E_{r+1}$ subalgebra chains and their string interpretation}
We can now inspect the algebraic properties of the solvable Lie algebras
$Solv_{r+1}$ defined by eq. \eqn{maxsolv1} and illustrate the match between
these properties and the physical properties of the sequential compactification.
\par
Due to the specific structure \eqn{maxsolv1} of a maximal rank solvable Lie algebra
every chain of {\it regular embeddings}:
\begin{equation}
E_{r+1} \, \supset \,K^{0}_{r+1} \, \supset \, K^{1}_{r+1}\, \supset \, \dots \, \supset \,
 K^{i}_{r+1}\, \supset \, \dots
\label{aletto1}
\end{equation}
where $K^{i}_{r+1}$ are subalgebras of the same rank and with
the same Cartan subalgebra ${\cal H}_{r+1}$ as
$E_{r+1}$ reflects into  a corresponding sequence of embeddings
of solvable Lie algebras and,
 henceforth, of  homogenous non--compact scalar manifolds:
\begin{equation}
E_{r+1}/H_{r+1} \, \supset \,K^{0}_{r+1}/Q^{0}_{r+1} \,\supset \,  \dots  \,\supset \,
K^{i}_{r+1}/Q^{i}_{r+1}
\label{caten1}
\end{equation}
which must be endowed with a physical interpretation.
In particular we can consider embedding chains such that \cite{witten}:
\begin{equation}
K^{i}_{r+1}= K^{i}_{r} \oplus X^{i}_{1}
\label{spacco}
\end{equation}
where $K^{i}_{r}$ is a regular subalgebra of $rank= r$ and $X^{i}_{1}$
is a regular subalgebra of rank one.
Because of the relation between the rank and the number
of compactified dimensions such chains clearly
correspond to the sequential dimensional reduction of either typeIIA (or B) or of M--theory.
Indeed the first of such regular embedding chains we can consider is:
\begin{equation}
K^{i}_{r+1}=E_{r+1-i}\, \oplus_{j=1}^{i} \, O(1,1)_j
\label{caten2}
\end{equation}
This chain simply tells us that the scalar manifold of
supergravity in dimension $D=10-r$ contains the
direct product of the supergravity scalar manifold  in dimension $D=10-r+1$
with the 1--dimensional moduli
space of a $1$--torus (i.e. the additional compactification radius one gets by making a further
step down in compactification).
\par
There are however additional embedding chains that originate from the different choices
of maximal
ordinary subalgebras admitted by the exceptional Lie algebra of the $E_{r+1}$ series.
\par
All the $E_{r+1}$ Lie algebras contain a subalgebra $D_{r}\oplus O(1,1)$ so
that we can write the chain \cite{noialtri}:
\begin{equation}
K^{i}_{r+1}=D_{r-i}\, \oplus_{j=1}^{i+1} \, O(1,1)_j
\label{dueachain}
\end{equation}
As we discuss more extensively in the subsequent two sections, and we already anticipated,
 the embedding chain \eqn{dueachain}
corresponds to the decomposition of the scalar manifolds into submanifolds spanned by either
 N-S or  R-R fields, keeping moreover track of the way they originate at each level of the
sequential dimensional reduction. Indeed the N--S fields correspond to generators of the
solvable Lie algebra that behave as integer (bosonic) representations of the
\begin{equation}
D_{r-i} \, \equiv \, SO(r-i,r-i)
\label{subalD}
\end{equation}
while R--R fields correspond to generators of the solvable Lie algebra assigned to the spinorial
representation of the subalgebras \eqn{subalD}.
A third chain of subalgebras is the following one:
\begin{equation}
K^{i}_{r+1}=A_{r-1-i}\,\oplus  \, A_1 \, \oplus_{j=1}^{i+1} \, O(1,1)_j
\label{duebchain}
\end{equation}
and a fourth one is
\begin{equation}
K^{i}_{r+1}=A_{r-i}\,  \oplus_{j=1}^{i+1} \, O(1,1)_j
\label{elechain}
\end{equation}
The physical interpretation of the \eqn{duebchain}, illustrated in the next subsection, has its
origin in type IIB string theory. The same supergravity effective lagrangian can be viewed as
the result of compactifying either version of type II string theory. If we take the IIB
interpretation
the distinctive fact is that there is, already at the $10$--dimensional level a complex scalar
field $\Sigma$ spanning the non--compact coset manifold $SL(2,\IR)_U/O(2)$.
The $10$--dimensional U--duality
group  $SL(2,\IR)_U$ must therefore be present in all lower dimensions and it
corresponds to the addend
$A_1$ of the chain \eqn{duebchain}.
\par
The fourth chain \eqn{elechain} has its origin  in an M--theory interpretation or in a
 physical problem posed by the
$D=4$ theory.
\par
If we compactify the $D=11$ M--theory to $D=10-r$ dimensions using an $(r+1)$--torus $T_{r+1}$,
the flat metric on this is parametrized by the coset manifold $GL(r+1) / O(r+1)$.
The isometry group of the $(r+1)$--torus moduli space is therefore $GL(r+1)$ and its
Lie Algebra is $A_r + O(1,1)$, explaining the chain \eqn{elechain}.
Alternatively, we may consider the origin of the same chain from a $D=4$ viewpoint.
There
 the electric vector field strengths do not span an irreducible representation
of the U--duality group $E_7$ but sit together with their magnetic counterparts in the irreducible
fundamental ${\bf 56}$ representation.  An important question therefore is that of
establishing which subgroup $G_{el}\subset E_7$ has an electric action on the field strengths. The
answer is \cite{hull}:
\begin{equation}
G_{el} \, = \, SL(8, \IR )
\end{equation}
since it is precisely with respect to this subgroup that the fundamental ${\bf 56}$ representation
of $E_7$
splits into: ${\bf 56}= {\bf 28}\oplus {\bf 28}$. The Lie algebra of the electric subgroup is
$A_7 \, \subset \, E_7$ and it contains an obvious subalgebra $A_6 \oplus O(1,1)$.
The intersection
of this latter with the subalgebra chain \eqn{caten2} produces the electric chain \eqn{elechain}.
In other words, by means of equation \eqn{elechain} we can trace back in each upper dimension
which
symmetries will maintain an electric action also at the end point of the dimensional reduction
sequence,
namely also in $D=4$.
\par
We have  spelled out the embedding chains of subalgebras that are physically significant from
a string theory viewpoint. The natural question to pose now  is  how to understand their
algebraic
origin and how to encode them in an efficient description holding true sequentially in all
dimensions,
namely for all choices of the rank $r+1=7,6,5,4,3,2$. The answer is provided by reviewing the
explicit construction of the $E_{r+1}$ root spaces in terms of $r+1$--dimensional
euclidean vectors
\cite{gilmore}.
\subsection{Structure of the  $E_{r+1(r+1)}$ root spaces and of the
associated solvable algebras}
The root system  of type $E_{r+1(r+1)}$  can be described
for all values of $1\le r \le 6$ in the following way. As any other
root system it is a finite subset of vectors $\Phi_{r+1}\, \subset\, \IR^{r+1}$
such that $\forall \alpha ,\beta \, \in \Phi_{r+1}$ one has
$ \langle \alpha , \beta \rangle \, \equiv  2 (\alpha , \beta )/ (\alpha , \alpha) \,
\in \, \ZZ $ and such that $\Phi_{r+1}$ is invariant with respect to
the reflections generated by any of its elements.
\par
\vskip 0.2cm
The root system is
given by the following set of length 2 vectors:
\par
\leftline{ \underline {\sl For $2 \le r \le 5$}}
\begin{equation}
 \Phi_{r+1} \, = \,  \left \{
\matrix { \mbox{roots} & \mbox{number} \cr
\null & \null \cr
{\underbrace {\quad \pm \, \epsilon_k \,\quad \pm \, \epsilon_\ell \quad}} & 4 \times
\left (\matrix { r \cr 2} \right )\cr
1 \,\le \,  k \, < \, \ell \, \le r  & \null \cr
\null & \null \cr
 {
\frac 1 2 \, \left ( \pm\epsilon_1 \pm \epsilon_2 \pm \dots \epsilon_r
\right ) \, \pm \, \sqrt{2 - \frac r 4 } \, \epsilon_{r+1} }  &   2^r \cr
 \null & \null \cr
}    \right \}
\label{erodd}
\end{equation}
\vskip 0.2cm
\par
\leftline{ \underline {\sl For $r=6$  }}
\begin{equation}
 \Phi_{7} \, = \,  \left \{
\matrix { \mbox{roots} & \null &\mbox{number} \cr
\null & \null & \null \cr
{\underbrace {\quad \pm \, \epsilon_k \,\quad \pm \, \epsilon_\ell \quad}} & \null & 60
 \cr
1 \,\le \,  k \, < \, \ell \, \le 6  & \null & \null \cr
\null & \null \cr
 \pm \, \sqrt {2} \epsilon_7 & \null & 2 \cr
 \null & \null & \null\cr
 {\underbrace {
\frac 1 2 \, \left (\pm \epsilon_1 \pm \epsilon_2 \pm \dots \epsilon_6
\right ) }}& \pm \, \sqrt{2 - \frac 3 2 } \, \epsilon_{7}   &   64 \cr
\mbox{even number of + signs} & \null & \null \cr
 \null & \null & \null \cr
}    \right \}
\label{ersix}
\end{equation}
where  $\epsilon_i$ ($i=1,\dots , r+1)$ denote a complete set
of orthonormal vectors.
 As far as the roots of the form
$(1/2)( \pm\epsilon_1 \pm \epsilon_2 \pm \dots \epsilon_r ) \,
\pm \, \sqrt{2 - (r/4) } \, \epsilon_{r+1}$ in \eqn{erodd} are concerned,
the following conditions on the number of plus signs in their expression are
 understood: in the case {\it r=even} the number of plus signs within the
round brackets must be even, while in the case {\it r=odd} there must be an
 overall even number of plus signs. These conditions are implicit also in
\eqn{eroddsol}.
The {\it r$=1$} case is degenerate for $\Phi_2$ consists of the only
roots $\pm [(1/2)\epsilon_1+\sqrt{7}/2 \epsilon_2]$.
\par
For all values of $r$ one can find a set of simple roots
$\alpha_1 , \alpha_2 . \dots \, \alpha_{r+1}$ such that the corresponding
Dynkin diagram is the standard one given in figure \eqn{standar}
\iffigs
\begin{figure}
\caption{}
\label{standar}
\epsfxsize = 10cm
\epsffile{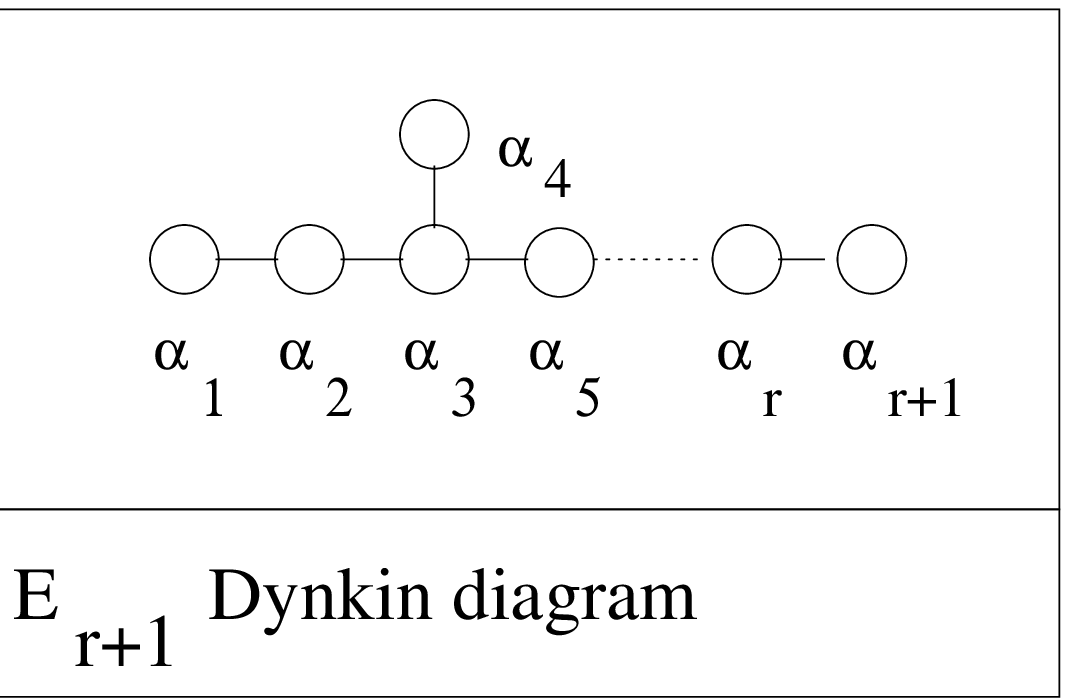}
\vskip -0.1cm
\unitlength=1mm
\end{figure}
\fi
\par
Consequently we can explicitly list the generators of all the
relevant solvable algebras for $r=1,\dots,6$ as follows:
\par
\leftline{ \underline {\sl For $2 \le r \le 5$ }}
\begin{eqnarray}
  & Solv_{r+1} \, = \, & \nonumber \\
  & \left \{
\matrix { \null & \mbox{number} & \mbox{type}\cr
\null & \null & \null \cr
\mbox{Cartan gener.} & r+1 & \mbox{NS} \cr
\null & \null & \null \cr
\mbox{roots} & \null  & \null \cr
\null & \null & \null \cr
{\underbrace {\quad  \, \epsilon_k \,\quad \pm \, \epsilon_\ell \quad}} & 2 \times
\left (\matrix { r \cr 2} \right ) & \mbox{NS} \cr
1 \,\le \,  k \, < \, \ell \, \le r  & \null & \null \cr
\null & \null & \null \cr
 {
\frac 1 2 \, \left (\pm \epsilon_1 \pm \epsilon_2 \pm \dots \epsilon_r
\right ) \, + \, \sqrt{2 - \frac r 4 } \, \epsilon_{r+1} }  &   2^{r-1} & \mbox{RR} \cr
\null & \null & \null \cr
 \null  & 2^{r-1}+r^2+1 &  \mbox{= Total } \cr
}    \right \} & \nonumber\\
\label{eroddsol}
\end{eqnarray}
\vskip 0.2cm
\par
\leftline{ \underline {\sl For $r=6$}}
\begin{eqnarray}
 & Solv_{7} \, = \, & \nonumber\\
 & \left \{
\matrix { \null & \null &\mbox{number} & \mbox{type}\cr
\null & \null & \null & \null\cr
\mbox{Cartan gener.} & \null & 7 & \mbox{NS} \cr
\null & \null & \null & \null \cr
\mbox{roots} & \null & \null & \null \cr
\null & \null & \null & \null \cr
{\underbrace {\quad  \, \epsilon_k \,\quad \pm \, \epsilon_\ell \quad}}& \null & 30
& \mbox{NS} \cr
1 \,\le \,  k \, < \, \ell \, \le 6  & \null  & \null & \null \cr
\null & \null & \null & \null\cr
\sqrt{2} \, \epsilon_7 & \null & 1 &  \mbox{NS} \cr
\null & \null & \null & \null\cr
 {\underbrace {
\frac 1 2 \, \left (\pm \epsilon_1 \pm \epsilon_2 \pm \dots \epsilon_6
\right )}} &   +   \sqrt{2 - \frac 3 2 } \, \epsilon_{7}    &   32 & \mbox{RR}\cr
\mbox{even number of + signs} & \null & \null & \null \cr
 \null & \null & \null & \null\cr
 \null & \null  & 70 &  \mbox{= Total } \cr
}    \right \}& \nonumber \\
\label{er7sol}
\end{eqnarray}
Comparing eq.s \eqn{eroddsol} and \eqn{tabu2} we realize that the match between
the physical and algebraic counting of scalar fields relies on the following numerical
identities, applying to the R--R and N--S sectors respectively:
\begin{eqnarray}
  RR &: & \left\{\matrix{2^{r-1} &=& r + {1\over 6} r(r-1) (r-2) & (r=2,3,4)\cr
2^{r-1} &=& 1 + r + {1\over 6} r(r-1) (r-2) & (r=5)\cr
2^{r-1} &=& r + r + {1\over 6} r(r-1) (r-2) & (r=6)\cr}\right. \\
NS &:& \left\{ \matrix{2 \pmatrix{r\cr 2 \cr}+ r + 1 & = & 1 + r^2 & (r=2,3,4,5)\cr
2 \pmatrix{r\cr 2 \cr}+ r + 1 + 1  & = &  2 + r^2  & (r=6)\cr} \right.
\end{eqnarray}
The physical interpretation of these identities from the string viewpoint is further discussed
in the next section.
\vskip 0.2cm
\subsection{Simple roots and Dynkin diagrams}
The most efficient way to deal simultaneously with all the above root systems and
see the emergence of the above mentioned embedding chains is to embed them in the
largest, namely in the $E_7$ root space. Hence the various root systems $E_{r+1}$
will be represented by appropriate subsets of the full set of $E_7$ roots. In this
fashion for all choices of $r$ the $E_{r+1}$ are anyhow represented by 7--components
Euclidean vectors of length 2.
\par
To see the $E_7$ structure we just need to choose, among the positive roots of
\eqn{er7sol}, a set of seven simple roots $\alpha_1 , \dots \, \alpha_7$ whose
scalar products are those predicted by the $E_7$ Dynkin diagram. The appropriate choice
is the following:
\begin{eqnarray}
\alpha_1 =\left \{-\frac {1}{2},-\frac {1}{2},-\frac {1}{2}, -\frac {1}{2}, -\frac {1}{2},
-\frac {1}{2}, \frac{1}{\sqrt{2}}\right \}\nonumber\\
\alpha_2 = \left \{ 0,0,0,0,1,1,0 \right \}\nonumber\\
\alpha_3 = \left \{ 0,0,0,1,-1,0,0 \right \}\nonumber\\
\alpha_4 = \left \{ 0,0,0,0,1,-1,0 \right \} \nonumber\\
\alpha_5 = \left \{ 0,0,1,-1,0,0,0 \right \} \nonumber\\
\alpha_6 = \left \{ 0,1,-1,0,0,0,0 \right \} \nonumber\\
\alpha_7 = \left \{ 1,-1,0,0,0,0,0 \right \} \nonumber\\
\label{e7simple}
\end{eqnarray}
The embedding of chain \eqn{caten2} is now easily described: by considering the subset of
$r$ simple roots
$\alpha_1 , \alpha_2 \, \dots \, \alpha_r$ we realize the Dynkin diagrams of type $E_{r+1}$.
Correspondingly,
the subset of all roots pertaining to the root system $\Phi(E_{r+1}) \, \subset \,
\Phi(E_7)$ is given by:
\begin{eqnarray}
x&=&6-r+1\nonumber\\
\Phi(E_{r+1}) & \equiv & \cases{ \pm \epsilon_i \pm \epsilon_j \quad \quad x \le i < j \le 7 \cr
 \pm \left [ \frac{1}{2}\, \left ( -\epsilon_1,-\epsilon_2,\dots \, \pm\epsilon_x \,
 \pm \epsilon_{x+1},\dots
, \pm \epsilon_6 \right )+{\sqrt{2}\over 2} \, \epsilon_7 \right ] \cr }
\nonumber\\
\label{erp1let}
\end{eqnarray}
At each step of the sequential embedding one  generator of the $r+1$--dimensional
Cartan subalgebra
${\cal H}_{r+1}$ becomes orthogonal to the roots of the subsystem
$\Phi(E_{r})\subset\Phi(E_{r+1})$,
while the remaining $r$ span the Cartan subalgebra of $E_{r}$. If we name $H_{i}$ ($i=1,\dots ,7$)
the original
orthonormal basis of Cartan generators for the $E_7$ algebra, the Cartan generators
that are orthogonal
to all the roots of the $\Phi(E_{r+1})$ root system
at level $r$ of the embedding chain are the following $6-r$:
\begin{equation}
X_{k} = \left (\frac{1}{\sqrt{2}}\, H_7 + \frac {1}{k} \sum_{i=1}^{k} \, H_i \right )
\quad \quad k=1,
\dots \, 6-r
\label{extracart}
\end{equation}
On the other hand a basis for the Cartan subalgebra of the $E_{r+1}$ algebra embedded in $E_7$ is
given by :
\begin{eqnarray}
Y_i  & = & H_{6-r+i} \quad i=1,\dots \, r-1 \nonumber\\
Y_r  & = & (-)^{6-r} \, H_6 \nonumber\\
Y_{r+1} &=& \frac {1}{8-r} \, \left ( \sqrt{2} H_7 \, - \, \sum_{i=1}^{6-r}\, H_i \right )
\label{incart}
\end{eqnarray}
In order to visualize the other chains of subalgebras it is convenient to make two observations.
The first is to note that the simple roots selected in eq. \eqn{e7simple} are of two types: six
of them have integer components and span the Dynkin diagram of a $D_6 \equiv SO(6,6)$ subalgebra,
while the seventh simple root has half integer components and it is actually a spinor weight
with respect to this subalgebra. This observation leads to the embedding chain \eqn{dueachain}.
Indeed it suffices to discard one by one the last simple root to see the embedding of the
$D_{r-1}$ Lie algebra into $D_{r}\subset E_{r+1}$. As discussed in the next section $D_{r}$
is the Lie algebra of the T--duality group in type IIA toroidally compactified string theory.
\par
The next  observation is that the $E_7$ root system contains an exceptional pair of
roots $\beta =\pm \sqrt{2} \epsilon_7$, which does not belong to any of the other $\Phi (E_r)$
root systems. Physically the origin of this exceptional pair is very clear. It is associated
with the axion field $B_{\mu\nu}$ which in $D=4$ and only in $D=4$ can be dualized to an
additional scalar field. This root has not been chosen to be a simple root in eq.\eqn{e7simple}
since it can be regarded as a composite root in the $\alpha_i$ basis. However we have the
possibility
of discarding either $\alpha_2$ or $\alpha_1$ or  $\alpha_4$ in favour of $\beta$ obtaining a new
basis for the $7$-dimensional euclidean space $\IR^7$. The three choices in this operation
lead to the three different Dynkin diagrams given in fig.s (\ref{stdual}) and (\ref{elecal}), corresponding to
the Lie Algebras:
\begin{equation}
 A_5 \oplus A_2\, , \quad   D_6\oplus A_1  \, , \quad
  A_7
\label{splatto}
\end{equation}

\iffigs
\begin{figure}
\caption{}
\label{stdual}
\epsfxsize = 10cm
\epsffile{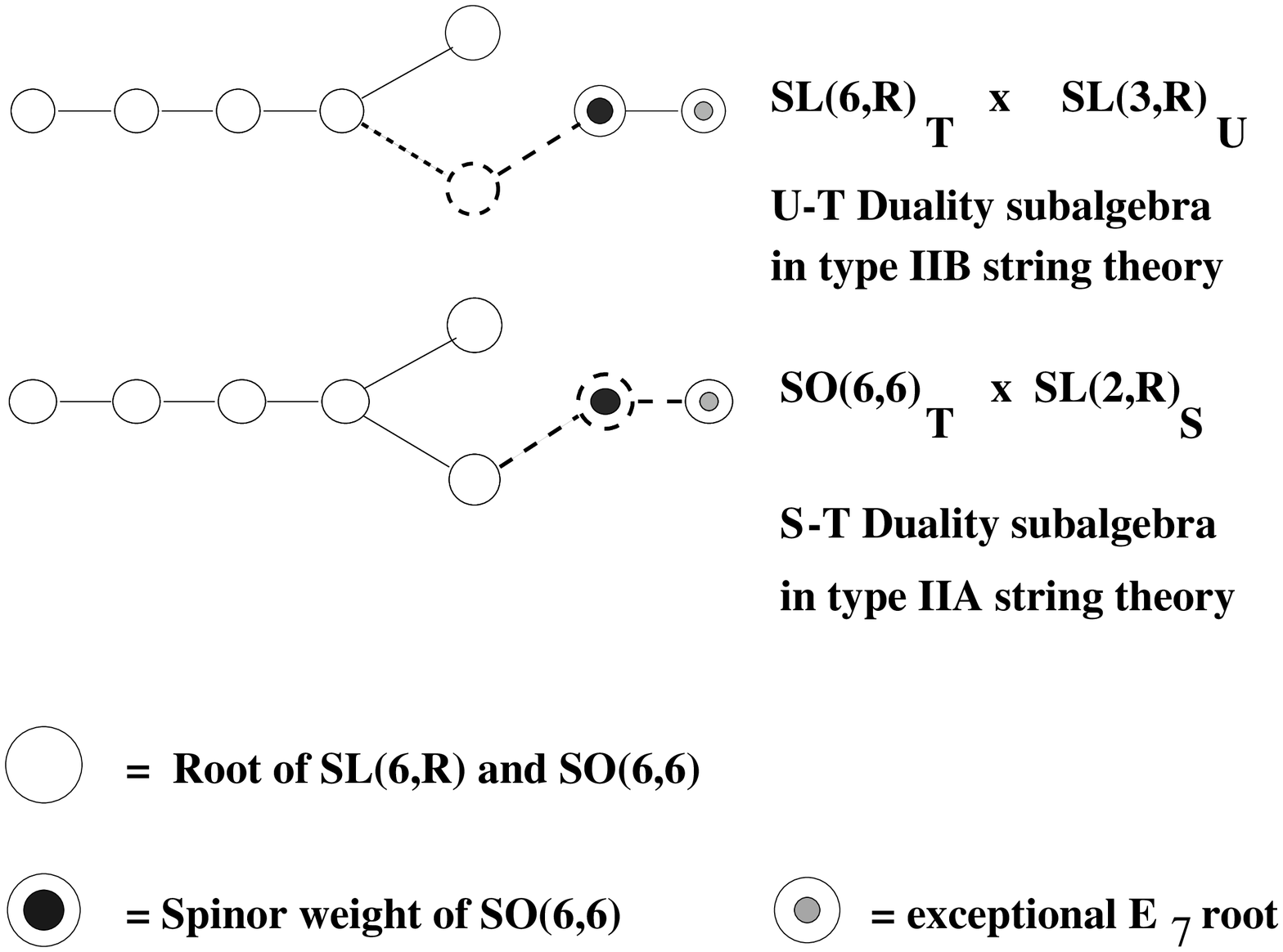}
\vskip -0.1cm
\unitlength=1mm
\end{figure}
\fi

\iffigs
\begin{figure}
\caption{}
\label{elecal}
\epsfxsize = 10cm
\epsffile{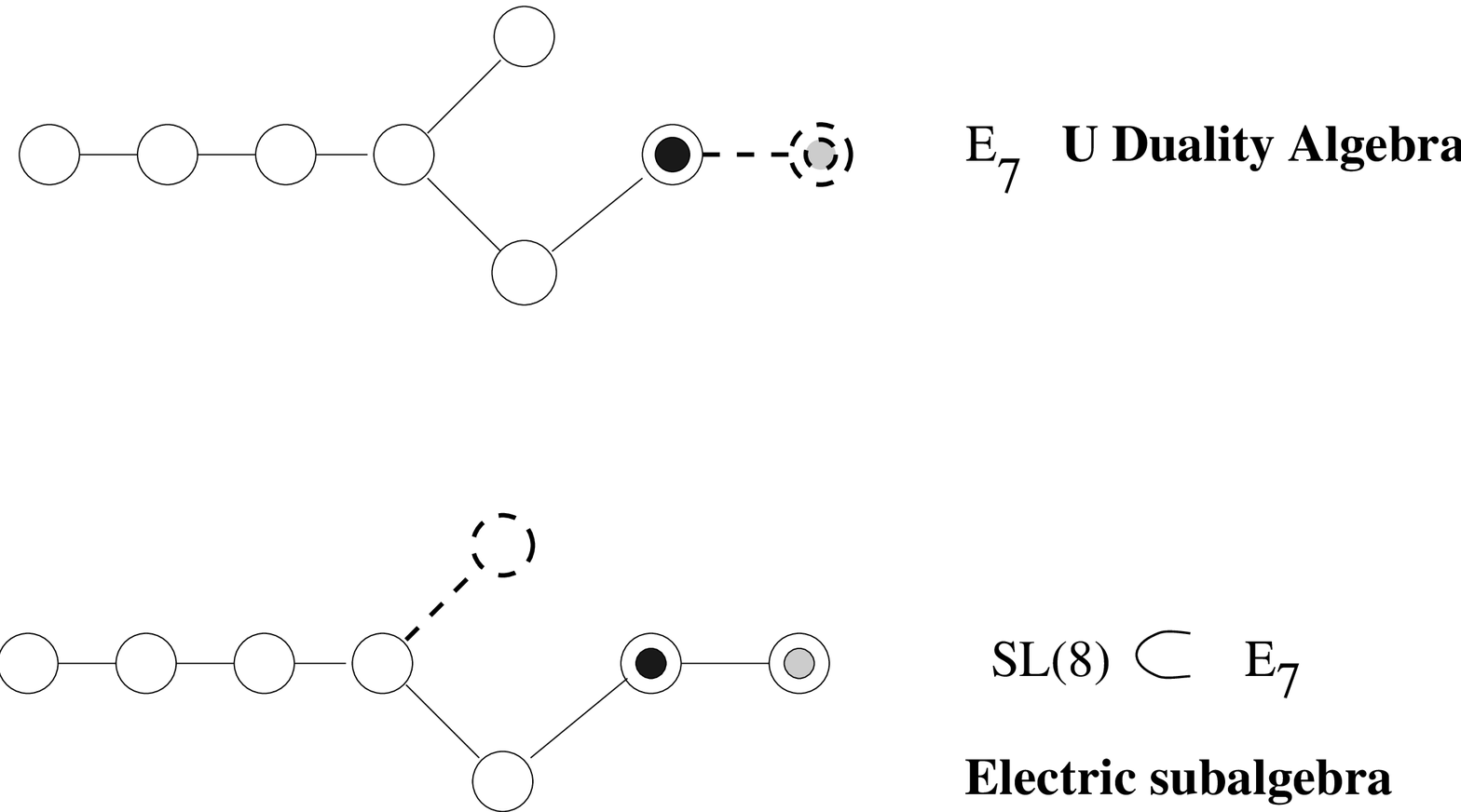}
\vskip -0.1cm
\unitlength=1mm
\end{figure}
\fi
From these embeddings occurring at the $E_7$ level, namely in $D=4$,
one deduces the three embedding chains
\eqn{dueachain},\eqn{duebchain},\eqn{elechain}: it just suffices to peal
off the last $\alpha_{r+1}$ roots
one by one and also the $\beta$ root that occurs only in $D=4$.
One observes that the appearance of the
$\beta$ root is always responsible for an enhancement of the S--duality group.
In the type IIA case
this group is enhanced from $O(1,1)$ to $SL(2,\IR)$ while in the type IIB case
it is enhanced from
the $SL(2,\IR)_U$ already existing in $10$--dimensions to $SL(3,\IR)$.
Physically this occurs by
combining the original dilaton field with the compactification radius of
the latest compactified
dimension.
\subsection{String theory interpretation of the sequential embeddings:
Type $IIA$, type $IIB$ and $M$ theory chains}
We now turn to a closer analysis of the physical meaning of
the embedding chains we have been illustrating.
\par
Let us begin with the chain of eq.(\eqn{duebchain})that, as anticipated, is
related with the type IIB interpretation of supergravity theory.
The distinctive feature of this chain of embeddings is the presence
of an addend $A_1$ that is already present in 10 dimensions. Indeed
this $A_1$ is the Lie algebra of the $SL(2,R)_\Sigma $ symmetry of type $IIB$
D=10 superstring. We can name this group the U--duality symmetry $U_{10}$ in
$D=10$. We can use the chain \eqn{duebchain} to trace it in lower dimensions.
Thus let us  consider the decomposition
\begin{eqnarray}
E_{r+1(r+1)} & \rightarrow & N_r \otimes SL(2,\IR) \nonumber\\
N_r & = &   A_{r-1} \otimes O(1,1)
\la{sl2r}
\end{eqnarray}
Obviously $N_r$ is not contained in the  $T$-duality group $O(r,r)$ since the
$NS$ tensor field $B_{\mu \nu}$ (which  mixes with the metric under
$T$-duality) and the $RR$--field $B^c_{\mu \nu}$ form a doublet with
respect $SL(2,\IR)_U$. In fact, $SL(2,\IR)_U$ and  $O(r,r)$ generate
the whole U--duality group $E_{r+1(r+1)}$. The appropriate interpretation
of the normaliser of $SL(2,R)_\Sigma$ in $E_{r+1(r+1)}$  is
\begin{equation}
 N_r  = O(1,1) \otimes SL(r,\IR) \equiv GL(r,\IR)
 \label{agnosco}
\end{equation}
where $GL(r,\IR)$ is the isometry group of the  classical moduli
space for the $T_r$ torus:
\begin{equation}
 \frac{GL(r,\IR)}{O(r)}.
\end{equation}
The decomposition of the U--duality group appropriate for the type $IIB$ theory is
\be
E_{r+1} \rightarrow U_{10} \otimes GL(r,\IR) = SL(2,\IR)_U \otimes O(1,1) \otimes SL(r,\IR).
\la{sl2rii}
\ee
Note that since $GL(r,\IR) \supset O(1,1)^r$, this translates into $E_{r+1} \supset
SL(2,\IR)_U \otimes O(1,1)^r$.  (In Type $IIA$, the corresponding chain would
be $E_{r+1} \supset O(1,1) \otimes O(r,r) \supset  O(1,1)^{r+1}$.)  Note that while
$SL(2,\IR)$ mixes $RR$ and $NS$ states, $GL(r,\IR)$ does not. Hence we can write the following
decomposition for the solvable Lie algebra:
\bea
Solv \left( \frac{E_{r+1}}{H_{r+1}} \right) &=& Solv \left(\frac{GL(r,\IR)}{O(r)} \otimes
\frac{SL(2,\IR)}{O(2)} \right) + \left(\frac{\bf r(r-1)}{\bf 2}, {\bf 2} \right) \oplus {\bf X}
\oplus {\bf Y}  \nonumber \\
\mbox{dim }Solv \left( \frac{E_{r+1}}{H_{r+1}} \right)&=& \frac{d(3d-1)}{2} + 2 + x + y.
\la{solvii}
\eea
where $x=\mbox{dim }{\bf X} $ counts the scalars coming from the internal part of the $4$--form
$A^+_{{ \mu}{ \nu}{ \rho}{\sigma}}$ of type IIB string theory.
We have:
\begin{equation}
x =  \left \{
\matrix { 0 & r<4 \cr
\frac{r!}{4!(r-4)!} & r\geq 4 \cr}\right.
\la{xscal}
\end{equation}
and
\begin{equation}
y =\mbox{dim }{\bf Y} = \cases{
\matrix { 0 &  r  < 6   \cr 2 &  r = 6  \cr}\cr}.
\la{yscal}
\end{equation}
counts the scalars arising from dualising the two-index tensor
fields in $r=6$.
\par
For example, consider the $ D=6$ case. Here the type $IIB$  decomposition is:
\begin{equation}
E_{5(5)}=\frac{O(5,5)}{O(5) \otimes O(5)} \rightarrow \frac{GL(4,\IR)}{O(4)}
\otimes \frac{SL(2,\IR)}{O(2)}
\la{exii}
\end{equation}
whose compact counterpart is given by $O(10) \rightarrow SU(4) \otimes SU(2) \otimes U(1)$,
corresponding to the decomposition: ${\bf 45} =
{\bf (15,1,1)}+ {\bf (1,3,1)} + {\bf (1,1,1)} +{\bf (6,2,2)} + {\bf (1,1,2)}$. It follows:
\be
Solv(\frac{E_{5(5)}}{O(5) \otimes O(5)}) = Solv (\frac{GL(4,\IR)}{O(4)}
\otimes \frac{SL(2,\IR)}{O(2)}) +({\bf 6},{\bf 2})^+ + ({\bf 1},{\bf 1})^+.
\la{solvexii}
\ee
where the factors on the right hand side parametrize the internal part of the metric $g_{ij}$,
the dilaton and the $RR$ scalar ($\phi$, $\phi^c$), ($B_{ij}$, $B^c_{ij}$) and $A^+_{ijkl}$
respectively.
\par
There is a connection between the decomposition (\eqn{sl2r}) and the corresponding chains
in M--theory. The type IIB chain is given by eq.(\eqn{duebchain}),
namely by
\begin{equation}
E_{r+1(r+1)} \rightarrow SL(2,\IR) \otimes GL(r,\IR)
\end{equation}
 while the $M$ theory is given by eq.(\eqn{elechain}), namely by
 \begin{equation}
E_{r+1} \rightarrow O(1,1) \otimes SL(r+1,\IR)
\end{equation}
coming from the moduli space of $T^{11-D} = T^{r+1}$.
We see that these decompositions involve the classical moduli spaces of $T^r$
 and of $T^{r+1}$ respectively.
Type $IIB$ and $M$ theory decompositions
become identical if we decompose further $SL(r, \IR) \rightarrow O(1,1)
\times SL(r-1,\IR)$ on the type $IIB$ side and  $SL(r+1, \IR) \rightarrow O(1,1)
\otimes SL(2,\IR) \otimes SL(r-1,\IR)$ on the $M$-theory side. Then we obtain for both theories
\be
E_{r+1} \rightarrow SL(2,\IR) \times O(1,1) \otimes O(1,1) \otimes SL(r-1,\IR),
\la{sl2rall}
\ee
and we see that the group $SL(2,\IR)_U$ of type $IIB$ is identified with the
complex structure of the $2$-torus factor of the total
compactification torus $T^{11-D} \rightarrow T^2 \otimes T^{9-D}$.
\par
Note that according to  \eqn{splatto} in 8 and 4 dimensions, ($r=2$ and $6$)
in the decomposition \eqn{sl2rall} there is the following enhancement:
\begin{eqnarray}
&  SL(2,\IR) \times O(1,1)  \rightarrow SL(3,\IR) \quad (\mbox{for} \, r=2,6) & \\
 &\left\{\matrix{O(1,1) & \rightarrow & SL(2,\IR) \quad (\mbox{for} \, r=2) \cr
SL(5,\IR) \times O(1,1) & \rightarrow & SL(6,\IR) \quad (\mbox{for} \, r=6) \cr}\right. &
\end{eqnarray}
Finally, by looking at fig.(\ref{wite5}) let us observe that
$E_{7(7)}$ admits also a subgroup $SL(2,\IR)_T$ $\otimes
(SO(5,5)_S$ $\equiv E_{5(5)})$ where the $SL(2,\IR)$ factor is a
T--duality group, while the factor $(SO(5,5)_S$ $\equiv E_{5(5)})$
is an S--duality group which mixes RR and NS states.
\iffigs
\begin{figure}
\caption{}
\label{wite5}
\epsfxsize = 10cm
\epsffile{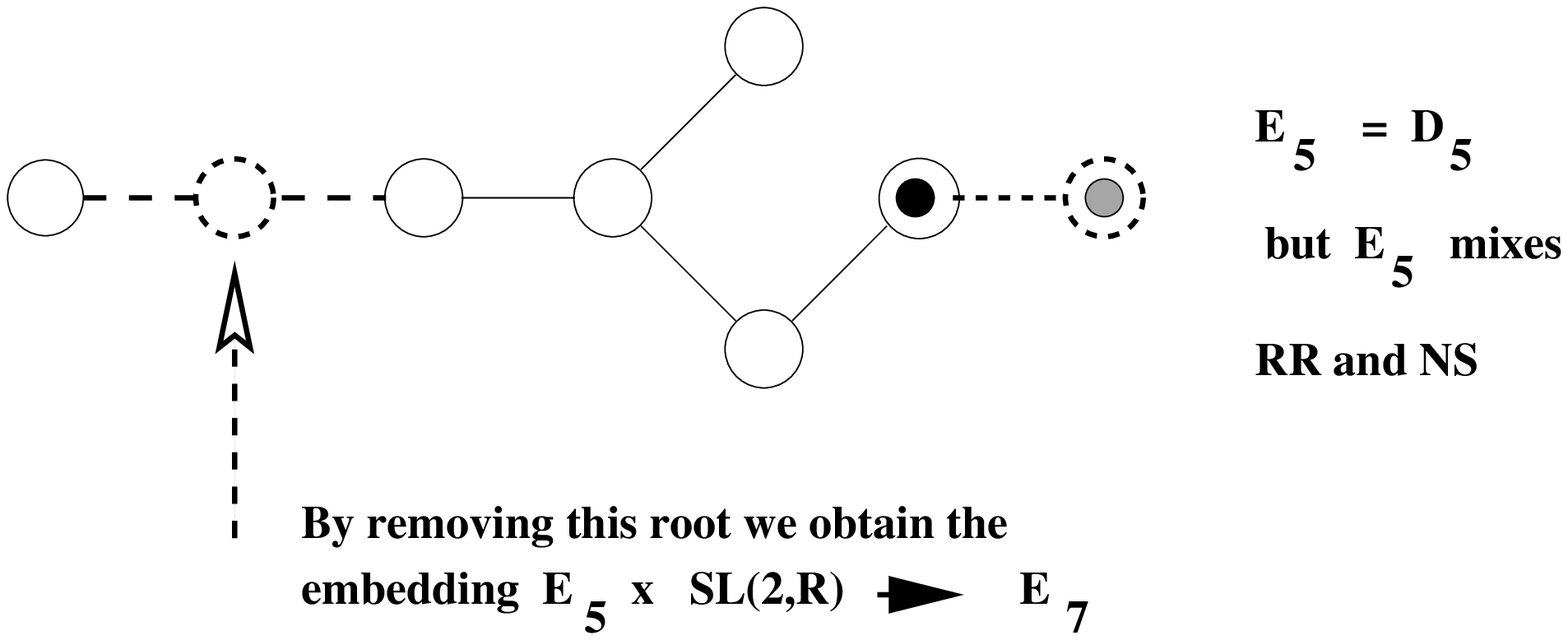}
\vskip -0.1cm
\unitlength=1mm
\end{figure}
\fi
\section{The maximal abelian ideals ${\cal A}_{r+1} \subset Solv_{r+1}$
of the solvable Lie algebra}

It is interesting to work out the maximal abelian ideals
${\cal A}_{r+1} \subset Solv_{r+1}$  of the
solvable Lie algebras generating the scalar manifolds
of maximal supergravity in dimension $D=10-r$.
The maximal abelian ideal of a solvable Lie
algebra is defined as the maximal subset of nilpotent generators  commuting among themselves.
From a physical point of view this is the largest abelian Lie algebra
that one might expect to be able to gauge in the supergravity theory. Indeed, as it turns
out, the number of vector fields in the theory is always larger
or equal than $\mbox{dim}{\cal A}_{r+1}$. Actually, as we are going to
see, the {\it gaugeable} maximal  abelian algebra
is always a proper subalgebra ${\cal A}^{gauge}_{r+1} \subset {\cal A}_{r+1}$
of this ideal.
\par
The criteria to determine ${\cal A}^{gauge}_{r+1}$ will be discussed
in the next section. In the present section we
derive ${\cal A}_{r+1}$ and we explore its relation with the
space of vector fields in one dimension above the dimension
we each time consider.  From such analysis we obtain a
filtration of the solvable Lie algebra which provides us
with a canonical polynomial parametrization of the supergravity
scalar coset manifold $U_{r+1}/H_{r+1}$
\par
\subsection{The maximal abelian ideal from an algebraic viewpoint}
Algebraically the maximal abelian ideal can be characterized by
looking at the decomposition of the U--duality algebra $E_{r+1(r+1)}$ with
respect to the U--duality algebra  in one dimension above.
In other words we have to consider the decomposition of $E_{r+1(r+1)}$ with
respect to the subalgebra $E_{r (r )} \, \otimes \, O(1,1)$. This
decomposition follows a general pattern which is given by the
next formula:
\begin{equation}
 \mbox{adj }  E_{r+1(r+1)} \, = \, \mbox{adj }  E_{r(r)}  \, \oplus
 \, \mbox{adj }  O(1,1) \, \oplus ( \ID^+_{r} \oplus \ID^-_{r} )
 \label{genpat}
\end{equation}
where $\ID^+_{r}$ is at the same time an irreducible representation of
the  U--duality algebra $E_{r (r )}$ in $D+1$ dimensions and coincides with the
maximal abelian ideal
\begin{equation}
\ID^+_{r}    \, \equiv \, {\cal A}_{r+1}   \, \subset \, Solv_{(r+1)}
\label{genmaxab}
\end{equation}
of the solvable  Lie algebra we are looking for. In eq. \eqn{genpat}
the subspace $\ID^-_{r}$ is just a second identical copy of the representation
 $\ID^{+}_{r}$ and it is made of negative rather than of positive weights
 of $E_{r (r )}$. Furthermore   $\ID^{+}_{r}$  and   $\ID^-_{r}$
 correspond to the eigenspaces belonging respectively to the eigenvalues
 $\pm 1$ with respect to the adjoint action of the S--duality group
 $O(1,1)$.
\subsection{The maximal abelian ideal from a physical perspective: the vector fields in one
dimension above and translational symmetries}
Here, we would like to show that the dimension of the abelian
ideal in $D$ dimensions is equal to the number of vectors in dimensions $D+1$.
Denoting the number of compactified dimensions by $r$ (in string theory,
$r=10-D$), we will label the $U$-duality group in $D$ dimensions by $U_D
= E_{11-D} = E_{r+1}$. The $T$-duality group is $O(r,r)$, while the
$S$-duality group is $O(1,1)$ in  dimensions higher than four, $SL(2,R)$
in $D=4$ (and it is inside $O(8,8)$ in $D=3$).
\par
It follows from \eqn{genpat}
that the total dimension of the abelian ideal is given by
\be
{\rm dim} \, \cA_{D} \,\equiv \,  {\rm dim} \, \cA_{r+1}
 \,\equiv \,  {\rm dim} \,
\ID_{r}
\la{abeliii}
\ee
where $\ID_{r} $ is a representation of $U_{D+1}$ pertaining to the vector fields.
 According to \eqn{genpat} we have (for $D \ge 4$):
\be
\mbox{adj } U_D = \mbox{adj } U_{D+1} \oplus {\bf 1} \oplus ({\bf 2}, \ID_r).
\la{irrepu}
\ee
This is just an immediate consequence of the embedding chain \eqn{caten2}
which at the first level of iteration yields
$E_{r+1} \rightarrow E_r \times O(1,1)$. For example, under
$E_7 \rightarrow E_6 \times O(1,1)$ we have the branching rule:
${\rm adj} \, E_7 = {\rm adj} \, E_6 + {\bf 1} + ({\bf 2},{\bf 27})$ and
the abelian ideal is given by the ${\bf 27^+}$ representation of the $E_{6(6)}$ group.
The $70$ scalars of the $D=4, N=8$ theory are
naturally decomposed as ${\bf 70} = {\bf 42} +{\bf 1} +{\bf 27^+}$.
To see the splitting of
the abelian ideal scalars into $NS$  and $RR$ sectors, one has to consider
the decomposition of $U_{D+1}$ under the T--duality group $T_{D+1} = O(r-1, r-1)$,
namely the second iteration of the embedding chain \eqn{caten2}: $E_{r+1}
\rightarrow O(1,1) \times O(r-1,r-1)$. Then the vector  representation of $O(r-1, r-1)$
gives  the $NS$ sector, while the  spinor representation yields  the $RR$
sector. The  example of $E_7$ considered above is somewhat exceptional, since
we have ${\bf 27} \rightarrow ({\bf 10} + {\bf 1} +{\bf 16})$.
Here in addition to the expected ${\bf 10}$ and
${\bf 16}$ of $O(5,5)$ we find an extra $NS$ scalar:  physically this is due
to the fact that in
four dimensions the two-index antisymmetric tensor field  $B_{\mu \nu}$
is dual to a scalar, algebraically this  generator is associated with the
exceptional root $ \sqrt {2} \epsilon_7$.
To summarize, the $NS$ and $RR$ sectors are separately  invariant under
$O(r,r)$ in $D=10-r$ dimensions, while the abelian $NS$  and $RR$ sectors
are invariant under $O(r-1, r-1)$. The standard parametrization of
the
$U_D/H_D$ and $U_{D+1}/H_{D+1}$ cosets gives a clear illustration of this
fact:
\be
\frac{U_D}{H_D} \sim ( \frac{U_{D+1}}{H_{D+1}}, r_{D+1}, {\bf V}_r^{D+1}).
\la{cosetideal}
\ee
Here $r_{D+1}$ stands for the compactification radius, and ${\bf V}_r^{D+1} $
are the compactified vectors yielding the abelian ideal in $D$ dimensions.
\par
Note that:
\begin{equation}
  \mbox{adj}\, H_D = \mbox{adj} \, H_{D+1} \, +\,
\mbox{adj} \,\mbox{Irrep} \, U_{D+1}
\end{equation}
so it appears  that the abelian ideal forms a representation not only of
$U_{D+1}$ but also of the compact isotropy subgroup $H_{D+1}$ of the scalar coset
manifold.
\par
In the  above $r=6$ example we find
${\rm adj} \, SU(8) = {\rm adj} \, USp(8) \oplus {\bf 27^-}$,
$ \Longrightarrow $ ${\bf 63}= {\bf 36}+ {\bf 27^-}$.
\subsection{Maximal abelian ideal and brane wrapping}
Now we would like to turn to a uniform counting  of the ideal dimension in
diverse space--time dimensions. The fact that the ($D+1$)-dimensional
vectors have $0$-branes as electric sources  (or equivalently,
 ($D-3$)-branes as magnetic ones) reduces the analysis of the $RR$ sector
to a simple exercise in counting the ways of wrapping  higher dimensional
$d$-branes around the cycles of the compact manifold.  This procedure spares
one from doing a case-by-case counting and worrying about the scalars
arising from the dualization of the tensor fields. It also easily
generalizes for manifolds other then $T^r$. The latter choice corresponds
to the case of maximal
preserved supersymmetry, for which the counting is presented here.
\par
Starting from Type $IIA$ theory with 0, 2, 4, 6 -Dbranes \cite{polc}, the total number of
($D+1$)-dimensional  0-Dbranes (i.e. the maximal abelian ideal in $D$--dimensions)
is  obtained
by wrapping the Dbranes around the even cycles of the ($9-D$)-dimensional torus.
One gets:
\be
n_{\cA}^{RR} = \sum_{k} b_{2k} (T^{9-D}) = 2^{8-D}
\la{rrideal}
\ee
where $b_{2k}$ are the Betti numbers. The same result is obtained by
counting the magnetic sources: in this case the sum is taken over
alternating series of even (odd) cohomology for $9-D$ even (odd), since
the $6$-Dbrane is wrapped on the  $(9-D)$-dimensional cycle of $T^{9-D}$,
the $4$-Dbrane on $(7-D)$-dimensional cycles and so on.  Note that  wrapping
 Dbranes around the cycles of the same dimensions as above but on a
$T^{10-D}$ yields the total number of the $RR$ scalars  in $D$ dimensions.
\par
The type $IIB$ story is exactly the same with the even cycles replaced by
the odd ones. The only little subtlety is in going from ten dimensions
to nine - there is no $0$-Dbrane in type $IIB$, but instead there is a
$RR$ scalar in the ideal
already in ten dimensions, since the $U$-duality group is non-trivial. Of course the
results agree on $T^r$ as they should on any manifold with a vanishing Euler number.
\par
The $NS$ parts of  the ideal ($(D-3)$-branes in ($D+1$)-dimensions) are
obtained either by  wrapping the ten-dimensional fivebrane or as magnetic
sources for Kaluza-Klein vectors (note that for the $NS$ part, the
reasonings for Type $IIA$ and $IIB$ are identical).  The former are the
fivebrane wrapped on $(8-D)$ cycles of $T^{9-D}$ (there are $(9-D)$ of
them), while the latter are given by the same number since it is the
number of the Kaluza--Klein vectors (number of $1$-cycles). Thus
\be
n_{\cA}^{NS} = 2 b_1 (T^{9-D}) = 2(9-D).
\la{nsideal}
\ee
The only exception to this formula is the $D=4$ case where, as discussed
above, we have to add an extra scalar due to the $B_{\mu \nu}$ field.

\section{Gauging}
In this last section we will consider the problem of gauging some isometries
of the coset $G/H$ in the framework of solvable Lie algebras.
\par
In particular we will consider in more detail the gauging of maximal compact groups
and the gauging of nilpotent abelian (translational) isometries.
\par
This procedure is a way of obtaining
partial supersymmetry breaking in extended supergravities \cite{hull},\cite{warn},\cite{huwa}
and it may find applications in the context of non perturbative phenomena in string
and M-theories.
\par
Let us consider the left--invariant 1--form $\Omega = L^{-1} dL$ of the coset
manifold $U_D/H_D$, where $L$ is the coset representative.
\par
The gauging procedure \cite{castdauriafre} amounts to the replacement of $dL$ with the gauge covariant
differential $\nabla L$ in the definition of the left--invariant 1--form $\Omega = L^{-1} d L$:
\begin{equation}
  \label{gaugedconn}
  \Omega  \rightarrow  \hat\Omega = L^{-1} \nabla  L =  L^{-1}( d + A )  L
= \Omega + L^{-1} A L
\end{equation}
As a consequence $\hat\Omega $ is no more a flat connection, but its curvature is given by:
\begin{equation}
  \label{gaugedchiusa}
 R(\hat\Omega)= d \hat\Omega +  \hat\Omega \wedge  \hat\Omega =  L^{-1}\cF  L \equiv  L^{-1}( dA + A\wedge A )  L
= L^{-1}(F^I T_I +L^I_{AB}T_I \bar\psi^A \psi^B )  L
\end{equation}
where $F^I$ is the gauged  supercovariant 2--form and $T_I$ are the generators of the gauge group
 embedded in the U--duality representation  of the vector fields.

 Indeed, by very definition, under the
full group $E_{r+1(r+1)}$ the gauge vectors are contained
in the representation $\ID_{r+1}$.
Yet, with respect to the gauge subgroup they must
transform in the adjoint representation, so that $\cG_D $ has to be chosen in such a way that:
\begin{equation}
\ID_{r+1} \, \stackrel{\cG_D}{\longrightarrow} \,\mbox{adj}\,\cG_D \, \oplus \, \mbox{rep}\,\cG_D
\label{brancione}
\end{equation}
where $\mbox{rep}\,\cG_D $ is some other representation of $\cG_D$ contained in the
above decomposition.

It is important to remark that vectors which are in $\mbox{rep} \cG_D$ (i.e. vectors which do not gauge $\cG_D$)
 may be required, by consistence
of the theory \cite{topiva}, to appear through their duals $(D-3)$--forms, as for instance happens
for $D=5$ \cite{gurowa}.
In an analogous way $p$--form potentials ($p\neq 1$) which are in non trivial representations
of $\cG_D$ may also be required to appear through their duals $(D-p-2)$--potentials,
as is the case in $D=7$ for $p=2$ \cite{pepiva}.

The charges and the boosted structure constants discussed in the next subsection can be retrieved from the two terms
appearing in the last expression of eq. \eqn{gaugedchiusa}

\subsection{Filtration of the $E_{r+1}$ root space, canonical
parametrization of the coset representatives and boosted structure constants}
As it has already been emphasized  in the introduction, the complete
structure of $N > 2$ supergravity in diverse dimensions is fully encoded in
the local differential geometry of the scalar coset manifold
$U_D/H_D$. All the couplings in the Lagrangian are described in terms
of the metric, the connection and the coset representative \eqn{cosrep1} of
$U_D/H_D$. A particularly significant consequence of extended
supersymmetry is that the fermion masses and the scalar potential the
theory can develop occur only as a consequence of the gauging and can
be extracted from a decomposition in terms of irreducible $H_D$ representations
of the {\sl boosted structure constants}\cite{mainaetal} \cite{castdauriafre}. Let us define these latter.
Let $\ID_{r+1}$ be the irreducible representation of the $U_D$
U--duality group pertaining to the vector fields and denote by ${\vec
{\bf w}}_\Lambda$ a basis for $\ID_{r+1}$:
\begin{equation}
\forall {\vec {\bf v}} \, \in \, \ID_{r+1}  \qquad : \qquad  {\vec v}\, = \,
v^\Lambda \, {\vec {\bf w}}_\Lambda
\label{baset}
\end{equation}
In the case we consider of maximal supergravity theories, where the
U--duality groups are given by  $E_{r+1(r+1)}$
 the basis vectors ${\vec {bf w}}_\Lambda$ can be identified
with the $56$ weights of the fundamental $E_{7(7)}$ representation or with
the subsets of this latter corresponding to the irreducible
representations of its  $E_{r+1(r+1)}$ subgroups, according to the
branching rules:
\begin{equation}
{\bf 56}{\stackrel{E_6}{\longrightarrow}}\cases{{\bf 27 + 1}{\stackrel{E_5}{\longrightarrow}}
 \cases{{\bf 16}{\stackrel{E_4}{\longrightarrow}}{\bf \dots}\cr
 {\bf 10}{\stackrel{E_4}{\longrightarrow}}{\bf \dots}\cr
 {\bf 1+1}{\stackrel{E_4}{\longrightarrow}}{\bf \dots}\cr} \cr
 {\bf 27 + 1}{\stackrel{E_5}{\longrightarrow}}
 \cases{{\bf 16}{\stackrel{E_4}{\longrightarrow}}{\bf \dots}\cr
 {\bf 10}{\stackrel{E_4}{\longrightarrow}}{\bf \dots}\cr
 {\bf 1+1}{\stackrel{E_4}{\longrightarrow}}{\bf \dots}\cr} \cr
 }
\label{brancio}
\end{equation}
Let:
\begin{equation}
< ~, ~ > \quad : \quad \ID_{r+1} \, \times \ID_{r+1} \, \longrightarrow \,
\IR
\label{norma}
\end{equation}
denote the invariant scalar product in $\ID_{r+1}$ and let
$ {\vec {\bf w}}^\Sigma$ be a dual basis such that
\begin{equation}
 < {\vec {\bf w}}^\Sigma , {\vec {\bf w}}_\Lambda > \, = \,
 \delta^\Sigma_\Lambda
 \label{dualba}
\end{equation}
Consider then the $\ID_{r+1}$ representation of the coset representative
\eqn{cosrep1}:
\begin{equation}
L(\phi) \quad : \quad  \vert {\vec {\bf w}}_\Lambda >  \, \longrightarrow
L(\phi)_\Lambda^\Sigma \, \vert {\vec {\bf w}}_\Sigma >,
\label{matricia}
\end{equation}
and let $T^I$ be the generators of the gauge algebra $\cG_D \,
\subset E_{r+1(r+1)}$.

  The only admitted generators are those with
index $\Lambda = I \, \in \, {adj }\,\cG_D $,
and there are no gauge group generators with index $\Lambda \, \in \,\mbox{rep}\,\cG_D  $.
Given these definitions
the {\it boosted structure constants} are the following three--linear
$3$--tensors in the coset representatives:
\begin{equation}
\IC_{\Sigma\Gamma}^{\Lambda}\left(\phi\right)\, \equiv \,
\sum _{I=1}^{{\rm dim }\cG_D} \,
< {\vec {\bf w}}^\Lambda \, , \, L^{-1}\left(\phi\right) \, T_I \,
  L\left(\phi\right)  \, {\vec {\bf w}}_\Sigma > \, < {\vec {\bf w}}^I \,
 , \, L\left(\phi\right) {\vec {\bf w}}_\Gamma >
 \label{busto}
\end{equation}
and by decomposing them into irreducible $H_{r+1}$ representations we
obtain the building blocks utilized by supergravity in the fermion
shifts, in the fermion mass--matrices and in the scalar potential.
\par
In an analogous way, the charges appearing in the gauged
covariant derivatives are given by the following general form:
\begin{equation}
Q_{I\Sigma}^{\Lambda} \, \equiv \,
< {\vec {\bf w}}^\Lambda \, , \, L^{-1}\left(\phi\right) \, T_I \,
  L\left(\phi\right)  \, {\vec {\bf w}}_\Sigma >
 \label{buscar}
\end{equation}
\par
The coset representative $L\left(\phi\right)$ can be written in a canonical polynomial
parametrization which should give a
simplifying  tool in
mastering the scalar field dependence of all  physical relevant
quantities.
This includes, besides mass matrices, fermion shifts and scalar potential,
also the central charges
 \cite{amicimiei}.
\par
The alluded parametrization is precisely what the solvable Lie
algebra analysis produces.
\par
To this effect let us decompose
the solvable Lie algebra of $E_{7(7)}/SU(8)$ in a sequential way utilizing
eq. \eqn{genpat}. Indeed we can write the equation:
\begin{equation}
Solv(E_{7(7)})={\cal H}_7 \oplus \Phi^{+}(E_{7})
\label{decompo1}
\end{equation}
where $\Phi^{+}(E_{7})$ is the $63$ dimensional positive part of the $E_7$ root space.
By repeatedly using eq.  \eqn{genpat} we obtain:
\begin{equation}
\Phi^{+}(E_7)=\Phi^+(E_2) \oplus \ID^{+}_{2} \oplus \ID^{+}_{3}
\oplus \ID^{+}_{4} \oplus \ID^{+}_{5}  \oplus \ID^{+}_{6}
\label{decompo}
\end{equation}
where $ \Phi^+(E_2) $ is the one--dimensional root space of the
U--duality group in $D=9$ and $\ID^{+}_{r+1}$ are the weight-spaces
of the $E_{r+1}$ irreducible
representations to which the vector field in $D=10-r$
are assigned. Alternatively, as we have already explained, ${\cal A}_{r+2}
\equiv \ID^{+}_{r+1}$ are
the maximal abelian ideals of the U--duality group in $E_{r+2}$ in
$D=10-r-1$ dimensions.
\par
We can easily check that the dimensions sum appropriately as follows from:
 \begin{eqnarray}
\mbox{dim}\, \Phi^{+}(E_7)\, &=& { 63} \nonumber\\
\mbox{dim}\, \Phi^+(E_2)\, &=& { 1}\quad\quad~\mbox{dim}\,
\ID^{+}_{2}\, = { 3}\nonumber\\
\mbox{dim}\, \ID^{+}_{3}\, &=& { 6}\quad\quad~\mbox{dim}\,
\ID^{+}_{4}\, = { 10}\nonumber\\
\mbox{dim}\, \ID^{+}_{5}\, &=& { 16}\quad\quad\mbox{dim}\,
\ID^{+}_{4}\, = { 27}\nonumber\\
\label{ideadim}
\end{eqnarray}
Relying on eq. \eqn{decompo1},
\eqn{decompo} we can introduce a canonical set of scalar field
variables:
\begin{eqnarray}
\phi^i & \longrightarrow &  Y_i \,\in {\cal H} \quad \quad i=1,\dots \,
r \nonumber\\
\tau^i_{k} & \longrightarrow &  \,D^{(k)}_{i} \, \in  \ID_k \quad i=1,\dots \,
\mbox{dim }\,\ID_k  \quad (k=2,\dots,6) \nonumber\\
\tau_1   & \longrightarrow & \ID_1 \, \equiv  \, E_2
\label{canoni}
\end{eqnarray}
and adopting the short hand notation:
\begin{eqnarray}
 \phi \cdot  {\cal H} & \equiv & \phi^i  \,   Y_i  \nonumber\\
 \tau_k \cdot \ID_k & \equiv & \tau^i_{k}  \,D^{(k)}_{i} \nonumber\\
 \label{fucili}
\end{eqnarray}
we can write the coset representative for maximal supergravity in dimension
$D=10-r$ as:
\begin{eqnarray}
L & = & \exp \left [ \phi \cdot  {\cal H} \right ] \,  \prod_{k=1}^{r}
\, \exp \left[   \tau_k \cdot \ID_k \right ] \nonumber\\
& = & \prod_{j=1}^{r+1} S^i \,  \prod_{k=1}^{r}   \,
\left( { 1} + \tau_r \cdot \ID_r \right)
\end{eqnarray}
The last line
follows from the abelian nature of the ideals $\ID_k$ and from the position:
\begin{equation}
S^i  \, \equiv \, \exp[\phi^i Y_i]
\end{equation}
All entries of the matrix $L$ are therefore polynomials of order at most  $2 \, r\, +\, 1$ in the
$S^i, \tau^i_k, \tau_1$ ``canonical'' variables.   Furthermore when the gauge group is chosen within the
maximal abelian ideal it is evident from the definition of the boosted
structure constants \eqn{busto} that they do not depend on the scalar
fields associated with the  generators of the same ideal. In such gauging one
has therefore {\it a flat direction} of the scalar potential for each
generator of the maximal abelian ideal.
\par
In the next section we turn to considering the possible gaugings more
closely.
\subsection{Gauging of compact and translational isometries}
A necessary condition for the gauging of a subgroup $\cG_D\subset U_D$ is that the representation of the vectors
$\ID_{r+1}$ must contain $\mbox{adj} \cG_D$.
Following this prescription, the list of maximal compact gaugings $\cG_D$ in any dimensions
 is obtained in  the third column of Table \ref{tabcomp}.
In the other columns we list the $U_D$-duality groups, their maximal compact subgroups
 and the left-over representations for vector fields.

\begin{table}[ht]
\caption{Maximal gauged compact groups}
\label{tabcomp}
\begin{center}
\begin{tabular}{|c|c|c|c|c|}
\hline
$D$ & $U_D$ & $H_D$ & $\cG_D$ & $\mbox{rep}\cG_D$ \\
\hline
\hline
9 & $SL(2,\IR) \times O(1,1)$ & $O(2)$ &  $O(2)$ & 2 \\
\hline
8 & $SL(3,\IR) \times SL(2,\IR)$ & $O(3) \times O(2)$   & $ O(3)$ & $3$  \\
\hline
7 & $SL(5,\IR)$ & $USp(4) $ &  $ O(5) \sim USp(4) $ & 0 \\
\hline
6 & $O(5,5)$ & $ USp(4) \times USp(4) $  &  $O(5)$ & $5+1$ \\
\hline
5 & $E_{6,(6)}$ & $USp(8)$ &  $ O(6) \sim SU(4) $ & $2\times 6$ \\
\hline
4 & $ E_{7(7)} $ & $SU(8)$ & $O(8)$ & 0 \\
\hline
\end{tabular}
\end{center}
\end{table}

\begin{table}[ht]
\caption{Transformation properties under $\cG_D$ of 2- and 3-forms}
 \label{tabrepba}
  \begin{center}
    \begin{tabular}{|c|c|c|}
\hline
$D$ &  $\mbox{rep} \, B_{\mu\nu}$  & $\mbox{rep} \, A_{\mu\nu\rho}$ \\
\hline
\hline
9 & 2 & 0 \\
\hline
8 & 3 & 0 \\
\hline
7 & 0 & 5 \\
\hline
6 & 5 & 0 \\
\hline
5 & $2\times 6$ & 0 \\
\hline
 \end{tabular}
  \end{center}
\end{table}

We notice that, for any $D$, there are $p$--forms ($p$ =1,2,3) which are charged under the gauge group $\cG_D$.
Consistency of these theories requires that such forms become massive.
It is worthwhile to mention how this can occur in two variants of the Higgs mechanism.
Let us define the (generalized) Higgs mechanism for a $p$--form  mass generation through the absorption
of a massless ($p-1$)--form (for $p=1$ this is the usual Higgs mechanism).
The first variant is the anti-Higgs mechanism for a $p$--form \cite{antihiggs},
which is its absorption by a massless ($p+1$)--form.
It is operating, for $p=1$, in $D=5,6,8,9$
for a sextet of $SU(4)$, a quintet of $SO(5)$, a triplet
of $SO(3)$ and a doublet of $SO(2)$, respectively.
The second variant is the self--Higgs mechanism \cite{topiva},
which only exists for $p = (D-1)/2$, $D= 4k-1$.
This is a massless $p$--form which acquires a mass through a
topological mass term and therefore it
becomes a massive ``chiral'' $p$--form.
The latter phenomena was shown to occur in $D=3$ and $7$.
It is amazing to notice that the representation assignments dictated by
$U$--duality for the various $p$--forms is
precisely that needed for consistency of the gauging procedure (see Table \ref{tabrepba}).
\par
The other compact gaugings listed in Table \ref{tabcomp} are the $D=4$
\cite{dwni} and $D=8$ cases \cite{sase2}.
\par
It is possible to extend the analysis of gauging semisimple groups
also to the case of solvable Lie groups \cite{fegipo}.
For the maximal abelian ideals of $Solv(U_D/H_D)$ this amounts
to gauge an $n$--dimensional subgroup of the translational
symmetries under which at least $n$ vectors are inert. Indeed the
vectors the set of vectors that can gauge an abelian algebra
(being in its adjoint representation) must be neutral under the
action of such an algebra.
We find that in  any dimension $D$ the dimension of this abelian
group $\mbox{dim} \cG_{abel}$ is given precisely by ${\rm dim} (\mbox{rep} \cG_D)$
which appear in the decomposition of $\ID_{r+1}$ under $O(r+1)$.
We must stress that this criterium gives a necessary but not sufficient
condition for the existence of
the gauging of an abelian isometry group,  consistent with supersymmetry.
\begin{table}[ht]
    \caption{Decomposition of fields in representations of the compact  group $\cG_D=O(11-D)$}
    \label{tabcompab}
    \begin{center}\begin{tabular}{|c|c|c|c|c|}
    \hline
    & vect. irrep & adj($O(11-D)$) & $\cA$  & $\mbox{dim} \cG_{abel}$  \\
    \hline
    $D = 9$ & $1+2$ & 1 & 1 & 1 \\
   \hline
     $D= 8$& $ 3+3$ & 3 & 3 & 3  \\ \hline
     $D=7$ & $6 + 4 $ & 6 & 6 & 4  \\
    \hline
     $D=6$ & $ 10+ 5 + 1 $ & 10 & 10 &  $5 + 1 $ \\
    \hline
     $D=5$ & $15+6+6$ & 15 & $ 15+1$ &   $6+6$  \\
     \hline
     $D=4$ & $(21+7)\times 2$ & 21 & $21 + 1 \times 6$ & 7 \\
     \hline
    \end{tabular}\end{center}\end{table}

\section*{Conclusions}
In this paper we have analyzed some properties of the effective  theories of type
II strings and M--theory in the framework of solvable Lie algebras.
\par
Particular attention has been given to the classification of R--R and N--S sectors as well
 as to the translational symmetries of the classical moduli space.
\par
The problem of gauging isometries in this framework has been reconsidered
with emphasis to its interplay with U--duality.
\par
It is hoped that further developments of these results may find applications in the study of non perturbative
dynamics of string theories, such as the finding of new vacua and the study of supersymmetry breaking.

\section*{ Appendix A: Explicit pairing between Lie algebra
roots and fields}

By referring to the toroidal dimensional reduction of type IIA
superstring and Tables \ref{tabu1}, \ref{tabu2},
it is straightforward to establish a correspondence
between the scalar fields of either Neveu--Schwarz or Ramond--Ramond type
emerging at each step of the sequential compactification and the positive
roots of $\Phi^{+}(E_7)$, distributed into
the  $\ID^{+}_{r+1}$ subspaces.
This gives rise to the following list of solvable algebra generators.
The roots of $SO(6,6)$ are associated with N--S fields,
the spinor weights of $SO(6,6)$ are associated with
R--R fields.
\par
\vskip 0.2cm
The abelian ideal in $D=8$   $  E_3 \supset{\cal A}_3 \equiv \ID^{+}_{2}$  is given
by the roots:
{\flushleft{$\ID^{+}_2 = $}}
\begin{eqnarray}
B_{9,10} ~\rightarrow ~  D_{2} (1) &=&  \{ 0,0,0,0,1,1,0\} \nonumber \\
g_{9,10} ~\rightarrow ~  D_{2} (2) &=& \{ 0,0,0,0,1,-1,0\} \nonumber \\
A_9 ~\rightarrow ~  D_{2} (3) &=&\{ -{1\over 2},-{1\over 2},-{1\over 2},
-{1\over 2},{1\over 2},{1\over 2},
   {1\over {{\sqrt{2}}}} \}
\label{d2root}
\end{eqnarray}
The abelian ideal in $D=7$   $  E_4\supset {\cal A}_4\equiv \ID^{+}_{3}$  is given
by the roots:
{\flushleft{$\ID^{+}_3 = $}}
\begin{eqnarray}
B_{8,9} ~\rightarrow ~  D_{3} (1) &=&  \{ 0,0,0,1,1,0,0\} \nonumber \\
g_{8,9} ~\rightarrow ~  D_{3} (2) &=&\{ 0,0,0,1,-1,0,0\} \nonumber\\
B_{8,10} ~\rightarrow ~  D_{3} (3) &=& \{ 0,0,0,1,0,1,0\}\nonumber \\
g_{8,10}~\rightarrow ~  D_{3} (4) &=&   \{ 0,0,0,1,0,-1,0\} \nonumber \\
A_8 ~\rightarrow ~  D_{3} (5) &=& \{ -{1\over 2},-{1\over 2},-{1\over 2},{1\over 2},
   {1\over 2},-{1\over 2},{1\over {{\sqrt{2}}}}\} \nonumber \\
A_{8,9,10} ~\rightarrow ~  D_{3} (6) &=&  \{ -{1\over 2},-{1\over 2},-{1\over 2},{1\over 2},
-{1\over 2},{1\over 2},
   {1\over {{\sqrt{2}}}}\}
\label{d3root}
\end{eqnarray}
The abelian ideal in $D=6$   $ E_5\supset  {\cal A}_5 \equiv \ID^{+}_{4}$  is given
by the roots:
{\flushleft{$\ID^{+}_4 = $} }
\begin{eqnarray}
B_{7,8} ~\rightarrow ~  D_{4} (1) &=&  \{ 0,0,1,1,0,0,0\}\nonumber \\
g_{7,8} ~\rightarrow ~  D_{4} (2) &=& \{ 0,0,1,-1,0,0,0\}\nonumber \\
B_{7,9} ~\rightarrow ~  D_{4} (3) &=&\{ 0,0,1,0,1,0,0\}\nonumber \\
g_{7,9} ~\rightarrow ~  D_{4} (4) &=&  \{ 0,0,1,0,-1,0,0\}\nonumber \\
B_{7,10} ~\rightarrow ~  D_{4} (5) &=& \{ 0,0,1,0,0,1,0\}\nonumber \\
g_{7,10} ~\rightarrow ~  D_{4} (6) &=& \{ 0,0,1,0,0,-1,0\}\nonumber \\
A_{7,9,10}~\rightarrow ~  D_{4} (7) &=&  \{ -{1\over 2},-{1\over 2},{1\over 2},
{1\over 2},-{1\over 2},-{1\over 2},
   {1\over {{\sqrt{2}}}}\}\nonumber \\
A_{7,8,10}~\rightarrow ~  D_{4} (8) &=&\{ -{1\over 2},-{1\over 2},{1\over 2},-{1\over 2},
   {1\over 2},-{1\over 2},{1\over {{\sqrt{2}}}}\}\nonumber \\
 A_{7,8,9}~\rightarrow ~   D_{4}(9)&=& \{ -{1\over 2},-{1\over 2},
 {1\over 2},-{1\over 2},-{1\over 2},{1\over 2},
   {1\over {{\sqrt{2}}}}\}\nonumber \\
A_{7} ~\rightarrow ~   D_{4}(10)&=&\{ -{1\over 2},-{1\over 2},{1\over 2},{1\over 2},
   {1\over 2},{1\over 2},{1\over {{\sqrt{2}}}} \}
\label{d4root}
\end{eqnarray}
The abelian ideal in $D=5$   $ E_6 \supset  {\cal A}_6\equiv \ID^{+}_{5}$  is given
by the roots:
{\flushleft{$\ID^{+}_5 = $}}
\begin{eqnarray}
B_{6,7}~\rightarrow ~   D_{5}(1)&=& \{ 0,1,1,0,0,0,0\} \nonumber \\
g_{6,7}~\rightarrow ~   D_{5}(2)&=& \{ 0,1,-1,0,0,0,0\} \nonumber \\
B_{6,8}~\rightarrow ~   D_{5}(3)&=& \{ 0,1,0,1,0,0,0\} \nonumber \\
g_{6,8}~\rightarrow ~   D_{5}(4)&=&  \{ 0,1,0,-1,0,0,0\} \nonumber \\
B_{6,9}~\rightarrow ~   D_{5}(5)&=& \{ 0,1,0,0,1,0,0\} \nonumber \\
g_{6,9}~\rightarrow ~   D_{5}(6)&=&\{ 0,1,0,0,-1,0,0\} \nonumber \\
B_{6,10}~\rightarrow ~   D_{5}(7)&=&  \{ 0,1,0,0,0,1,0\} \nonumber \\
g_{6,10}~\rightarrow ~   D_{5}(8)&=&\{ 0,1,0,0,0,-1,0\} \nonumber \\
A_{6,8,9}~\rightarrow ~   D_{5}(9)&=&  \{ -{1\over 2},{1\over 2},{1\over 2},-{1\over 2},
-{1\over 2},-{1\over 2},
   {1\over {{\sqrt{2}}}}\} \nonumber \\
A_{6,7,9}~\rightarrow ~   D_{5}(10)&=& \{ -{1\over 2},{1\over 2},-{1\over 2},{1\over 2},
   -{1\over 2},-{1\over 2},{1\over {{\sqrt{2}}}}\} \nonumber \\
A_{6,7,8}~\rightarrow ~   D_{5}(11)&=&  \{ -{1\over 2},{1\over 2},-{1\over 2},-{1\over 2},
{1\over 2},
-{1\over 2},
   {1\over {{\sqrt{2}}}}\} \nonumber \\
A_{\mu\nu\rho}~\rightarrow ~   D_{5}(12)&=& \{ -{1\over 2},{1\over 2},-{1\over 2},-{1\over 2},
   -{1\over 2},{1\over 2},{1\over {{\sqrt{2}}}}\} \nonumber \\
A_{6,7,10}~\rightarrow ~   D_{5}(13)&=& \{ -{1\over 2},{1\over 2},-{1\over 2},{1\over 2},
{1\over 2},{1\over 2},
   {1\over {{\sqrt{2}}}}\} \nonumber \\
A_{6,8,10}~\rightarrow ~   D_{5}(14)&=& \{ -{1\over 2},{1\over 2},{1\over 2},-{1\over 2},
   {1\over 2},{1\over 2},{1\over {{\sqrt{2}}}}\} \nonumber \\
A_{6,9,10}~\rightarrow ~   D_{5}(15)&=&  \{ -{1\over 2},{1\over 2},{1\over 2},{1\over 2},
-{1\over 2},{1\over 2},
   {1\over {{\sqrt{2}}}}\} \nonumber \\
A_6 ~\rightarrow ~   D_{5}(16)&=&\{ -{1\over 2},{1\over 2},{1\over 2},{1\over 2},
   {1\over 2},-{1\over 2},{1\over {{\sqrt{2}}}} \}
\label{d5root}
\end{eqnarray}
The abelian ideal in $D=4$   $  E_7\supset {\cal A}_7 \equiv \ID^{+}_{6}$  is given
by the roots:
{\flushleft{$\ID^{+}_6 = $}}
\begin{eqnarray}
B_{5,6}~\rightarrow ~   D_{6}(1)&=& \{ 1,1,0,0,0,0,0\} \nonumber \\
g_{5,6}~\rightarrow ~   D_{6}(2)&=& \{ 1,-1,0,0,0,0,0\} \nonumber \\
B_{5,7}~\rightarrow ~   D_{6}(3)&=& \{ 1,0,1,0,0,0,0\} \nonumber \\
g_{5,7}~\rightarrow ~   D_{6}(4)&=&  \{ 1,0,-1,0,0,0,0\} \nonumber \\
B_{5,8}~\rightarrow ~   D_{6}(5)&=& \{ 1,0,0,1,0,0,0\} \nonumber \\
g_{5,8}~\rightarrow ~   D_{6}(6)&=& \{ 1,0,0,-1,0,0,0\} \nonumber \\
B_{5,9}~\rightarrow ~   D_{6}(7)&=&  \{ 1,0,0,0,1,0,0\} \nonumber \\
g_{5,9}~\rightarrow ~   D_{6}(8)&=& \{ 1,0,0,0,-1,0,0\} \nonumber \\
B_{5,10}~\rightarrow ~   D_{6}(9)&=& \{ 1,0,0,0,0,0,1\} \nonumber \\
g_{5,10}~\rightarrow ~   D_{6}(10)&=& \{ 1,0,0,0,0,0,-1\} \nonumber \\
B_{\mu\nu}~\rightarrow ~   D_{6}(11)&=& \{ 0,0,0,0,0,0,{1\over {{\sqrt{2}}}}\} \nonumber \\
A_5 ~\rightarrow ~   D_{6}(12)&=& \{ {1\over 2},{1\over 2},{1\over 2},{1\over 2},{1\over 2},
{1\over 2},
   {1\over {{\sqrt{2}}}}\} \nonumber \\
A_{\mu\nu 6}~\rightarrow ~   D_{6}(13)&=& \{ {1\over 2},{1\over 2},-{1\over 2},-{1\over 2},
   -{1\over 2},-{1\over 2},{1\over {{\sqrt{2}}}}\} \nonumber \\
A_{\mu\nu 7} ~\rightarrow ~   D_{6}(14)&=&  \{ {1\over 2},-{1\over 2},{1\over 2},-{1\over 2},
-{1\over 2},-{1\over 2},
   {1\over {{\sqrt{2}}}}\} \nonumber \\
A_{\mu\nu 8} ~\rightarrow ~   D_{6}(15)&=& \{ {1\over 2},-{1\over 2},-{1\over 2},{1\over 2},
   -{1\over 2},-{1\over 2},{1\over {{\sqrt{2}}}}\} \nonumber \\
A_{\mu\nu 9} ~\rightarrow ~   D_{6}(16)&=&  \{ {1\over 2},-{1\over 2},-{1\over 2},-{1\over 2},
{1\over 2},-{1\over 2},
   {1\over {{\sqrt{2}}}}\} \nonumber \\
A_{\mu\nu 10} ~\rightarrow ~   D_{6}(17)&=& \{ {1\over 2},-{1\over 2},-{1\over 2},-{1\over 2},
   -{1\over 2},{1\over 2},{1\over {{\sqrt{2}}}}\} \nonumber \\
 A_{5,6,7} ~\rightarrow ~   D_{6}(18)&=& \{ {1\over 2},-{1\over 2},-{1\over 2},{1\over 2},
 {1\over 2},{1\over 2},
   {1\over {{\sqrt{2}}}}\} \nonumber \\
 A_{5,6,8} ~\rightarrow ~   D_{6}(19)&=&\{ {1\over 2},-{1\over 2},{1\over 2},-{1\over 2},
   {1\over 2},{1\over 2},{1\over {{\sqrt{2}}}}\} \nonumber \\
  A_{5,6,9} ~\rightarrow ~   D_{6}(20)&=& \{ {1\over 2},-{1\over 2},{1\over 2},{1\over 2},
  -{1\over 2},{1\over 2},
   {1\over {{\sqrt{2}}}}\} \nonumber \\
 A_{5,6,10} ~\rightarrow ~   D_{6}(21)&=&  \{ {1\over 2},-{1\over 2},{1\over 2},{1\over 2},
   {1\over 2},-{1\over 2},{1\over {{\sqrt{2}}}}\} \nonumber \\
  A_{5,7,8} ~\rightarrow ~   D_{6}(22)&=&  \{ {1\over 2},{1\over 2},-{1\over 2},-{1\over 2},
  {1\over 2},{1\over 2},
   {1\over {{\sqrt{2}}}}\} \nonumber \\
 A_{5,7,9} ~\rightarrow ~   D_{6}(23)&=&  \{ {1\over 2},{1\over 2},-{1\over 2},{1\over 2},
   -{1\over 2},{1\over 2},{1\over {{\sqrt{2}}}}\} \nonumber \\
 A_{5,7,10} ~\rightarrow ~   D_{6}(24)&=&   \{ {1\over 2},{1\over 2},-{1\over 2},{1\over 2},
 {1\over 2},-{1\over 2},
   {1\over {{\sqrt{2}}}}\} \nonumber \\
 A_{5,8,9} ~\rightarrow ~   D_{6}(25)&=& \{ {1\over 2},{1\over 2},{1\over 2},-{1\over 2},
   -{1\over 2},{1\over 2},{1\over {{\sqrt{2}}}}\} \nonumber \\
  A_{5,8,10} ~\rightarrow ~   D_{6}(26)&=&  \{ {1\over 2},{1\over 2},{1\over 2},-{1\over 2},{1\over 2},-{1\over 2},
   {1\over {{\sqrt{2}}}}\} \nonumber \\
  A_{5,9,10} ~\rightarrow ~   D_{6}(27)&=& \{ {1\over 2},{1\over 2},{1\over 2},{1\over 2},
   -{1\over 2},-{1\over 2},{1\over {{\sqrt{2}}}} \}
\label{d6root}
\end{eqnarray}
Finally, in $D=9$ we have the only root of the $E_2$ root space:
\begin{equation}
A_{10} ~\rightarrow ~ \Phi^+(E_2) \, = \,
\{ -{1\over 2},-{1\over 2},-{1\over 2},-{1\over 2},-{1\over 2},-{1\over 2},
  {1\over {{\sqrt{2}}}}\}
  \label{e2root}
\end{equation}
\section*{Appendix B: Representation matrices of the maximal abelian ideals}

In this appendix we list the matrices describing the action of the
abelian ideals on the space of vector fields. For all cases $D \ge
5$ the numbering of rows and columns of the matrix corresponds to the
listing of generators $D_r(i)$ given in the previous appendix. In the
case $D=4$ we need more care. The vector fields are associated with
a subset of $28$ weights of the $56$ fundamental weights of $E_7$.
For these weights we have chosen a conventional numbering that for
brevity we do not report in the present paper. Using this numbering
the following matrix describes the action of the 10 dimensional  subspace of $\ID^+_6$ made of
``electric'' generators (that is the intersection of the abelian ideal $\cA_7$ with
the ``electric'' subgroup $SL(8,\IR)$ of the U--duality group) on the 28 dimensional
column vector of the
 ``electric'' field strengths.
It is a linear combination $\sum s_i N_i$, where $N_i$ are the  ten nilpotent  generators
and $s_i$ the corresponding parameters of the solvable Lie algebra.
The maximal number of vector fields which correspond to gauging translational isometries
 is  found by looking at the maximal number of vectors which are annihilated by the maximal subset of abelian $N_i$
generators.
It turns out that in the present four dimensional case this number is 7.
{ \tiny
\begin{flushleft}
\begin{equation}
  \label{rapsei}
\left ( \matrix{ 0 & s1 & s2 & s3 & s4 & 0 & 0 & 0 & 0 & 0 & 0 & 0 & 0 & 0 & 0 & 0 & 0 & 0
   & 0 & 0 & 0 & 0 & 0 & 0 & 0 & 0 & 0 & 0 \cr 0 & 0 & 0 & 0 & 0 & 0 & 0 & 0
   & 0 & 0 & 0 & 0 & 0 & 0 & 0 & 0 & 0 & 0 & 0 & 0 & 0 & 0 & 0 & 0 & 0 & 0 & 0
   & 0 \cr 0 & 0 & 0 & 0 & 0 & 0 & 0 & 0 & 0 & 0 & 0 & 0 & 0 & 0 & 0 & 0 & 0
   & 0 & 0 & 0 & 0 & 0 & 0 & 0 & 0 & 0 & 0 & 0 \cr 0 & 0 & 0 & 0 & 0 & 0 & 0
   & 0 & 0 & 0 & 0 & 0 & 0 & 0 & 0 & 0 & 0 & 0 & 0 & 0 & 0 & 0 & 0 & 0 & 0 & 0
   & 0 & 0 \cr 0 & 0 & 0 & 0 & 0 & 0 & 0 & 0 & 0 & 0 & 0 & 0 & 0 & 0 & 0 & 0
   & 0 & 0 & 0 & 0 & 0 & 0 & 0 & 0 & 0 & 0 & 0 & 0 \cr 0 & 0 & 0 & 0 & 0 & 0
   & 0 & 0 & 0 & 0 & 0 & 0 & 0 & 0 & 0 & 0 & 0 & 0 & 0 & 0 & 0 & 0 & 0 & 0 & 0
   & 0 & 0 & 0 \cr 0 & 0 & 0 & 0 & 0 & 0 & 0 & s1 & s2 & s3 & s4 & 0 & s7 & s8 & s9
   & s10 & 0 & 0 & 0 & 0 & 0 & 0 & s5 & 0 & 0 & 0 & 0 & s6 \cr 0 & 0 & 0 & 0 & 0
   & 0 & 0 & 0 & 0 & 0 & 0 & 0 & 0 & 0 & 0 & 0 & s8 & s9 & s10 & 0 & 0 & 0 & 0 &
  5 & 0 & 0 & 0 & 0 \cr 0 & 0 & 0 & 0 & 0 & 0 & 0 & 0 & 0 & 0 & 0 & 0 & 0 & 0
   & 0 & 0 & s7 & 0 & 0 & s9 & s10 & 0 & 0 & 0 & s5 & 0 & 0 & 0 \cr 0 & 0 & 0 & 0
   & 0 & 0 & 0 & 0 & 0 & 0 & 0 & 0 & 0 & 0 & 0 & 0 & 0 & s7 & 0 & s8 & 0 & s10 &
  0 & 0 & 0 & s5 & 0 & 0 \cr 0 & 0 & 0 & 0 & 0 & 0 & 0 & 0 & 0 & 0 & 0 & 0 & 0
   & 0 & 0 & 0 & 0 & 0 & s7 & 0 & s8 & s9 & 0 & 0 & 0 & 0 & s5 & 0 \cr 0 & 0 & 0
   & 0 & 0 & 0 & 0 & 0 & 0 & 0 & 0 & 0 & 0 & 0 & 0 & 0 & 0 & 0 & 0 & 0 & 0 & 0
   & 0 & s7 & s8 & s9 & s10 & 0 \cr 0 & s6 & 0 & 0 & 0 & 0 & 0 & 0 & 0 & 0 & 0 & 0
   & 0 & 0 & 0 & 0 & s2 & s3 & s4 & 0 & 0 & 0 & 0 & 0 & 0 & 0 & 0 & 0 \cr 0 & 0
   & s6 & 0 & 0 & 0 & 0 & 0 & 0 & 0 & 0 & 0 & 0 & 0 & 0 & 0 & s1 & 0 & 0 & s3 & s4
   & 0 & 0 & 0 & 0 & 0 & 0 & 0 \cr 0 & 0 & 0 & s6 & 0 & 0 & 0 & 0 & 0 & 0 & 0
   & 0 & 0 & 0 & 0 & 0 & 0 & s1 & 0 & s2 & 0 & s4 & 0 & 0 & 0 & 0 & 0 & 0 \cr 0
   & 0 & 0 & 0 & s6 & 0 & 0 & 0 & 0 & 0 & 0 & 0 & 0 & 0 & 0 & 0 & 0 & 0 & s1 & 0
   & s2 & s3 & 0 & 0 & 0 & 0 & 0 & 0 \cr 0 & 0 & 0 & 0 & 0 & 0 & 0 & 0 & 0 & 0
   & 0 & 0 & 0 & 0 & 0 & 0 & 0 & 0 & 0 & 0 & 0 & 0 & 0 & 0 & 0 & 0 & 0 & 0
   \cr 0 & 0 & 0 & 0 & 0 & 0 & 0 & 0 & 0 & 0 & 0 & 0 & 0 & 0 & 0 & 0 & 0 & 0
   & 0 & 0 & 0 & 0 & 0 & 0 & 0 & 0 & 0 & 0 \cr 0 & 0 & 0 & 0 & 0 & 0 & 0 & 0
   & 0 & 0 & 0 & 0 & 0 & 0 & 0 & 0 & 0 & 0 & 0 & 0 & 0 & 0 & 0 & 0 & 0 & 0 & 0
   & 0 \cr 0 & 0 & 0 & 0 & 0 & 0 & 0 & 0 & 0 & 0 & 0 & 0 & 0 & 0 & 0 & 0 & 0
   & 0 & 0 & 0 & 0 & 0 & 0 & 0 & 0 & 0 & 0 & 0 \cr 0 & 0 & 0 & 0 & 0 & 0 & 0
   & 0 & 0 & 0 & 0 & 0 & 0 & 0 & 0 & 0 & 0 & 0 & 0 & 0 & 0 & 0 & 0 & 0 & 0 & 0
   & 0 & 0 \cr 0 & 0 & 0 & 0 & 0 & 0 & 0 & 0 & 0 & 0 & 0 & 0 & 0 & 0 & 0 & 0
   & 0 & 0 & 0 & 0 & 0 & 0 & 0 & 0 & 0 & 0 & 0 & 0 \cr 0 & 0 & 0 & 0 & 0 & s6
   & 0 & 0 & 0 & 0 & 0 & 0 & 0 & 0 & 0 & 0 & 0 & 0 & 0 & 0 & 0 & 0 & 0 & s1 & s2
   & s3 & s4 & 0 \cr 0 & 0 & 0 & 0 & 0 & 0 & 0 & 0 & 0 & 0 & 0 & 0 & 0 & 0 & 0
   & 0 & 0 & 0 & 0 & 0 & 0 & 0 & 0 & 0 & 0 & 0 & 0 & 0 \cr 0 & 0 & 0 & 0 & 0
   & 0 & 0 & 0 & 0 & 0 & 0 & 0 & 0 & 0 & 0 & 0 & 0 & 0 & 0 & 0 & 0 & 0 & 0 & 0
   & 0 & 0 & 0 & 0 \cr 0 & 0 & 0 & 0 & 0 & 0 & 0 & 0 & 0 & 0 & 0 & 0 & 0 & 0
   & 0 & 0 & 0 & 0 & 0 & 0 & 0 & 0 & 0 & 0 & 0 & 0 & 0 & 0 \cr 0 & 0 & 0 & 0
   & 0 & 0 & 0 & 0 & 0 & 0 & 0 & 0 & 0 & 0 & 0 & 0 & 0 & 0 & 0 & 0 & 0 & 0 & 0
   & 0 & 0 & 0 & 0 & 0 \cr 0 & s7 & s8 & s9 & s10 & s5 & 0 & 0 & 0 & 0 & 0 & 0 & 0
   & 0 & 0 & 0 & 0 & 0 & 0 & 0 & 0 & 0 & 0 & 0 & 0 & 0 & 0 & 0 \cr  }\right )
\end{equation}
\end{flushleft}
}
In $D=5$ the matrix $\sum s_i N_i$ is 27 dimensional and using
the numbering of eq.\eqn{d6root} is given by:
{\scriptsize
\begin{equation}
\label{rap5}
\pmatrix{ 0 & 0 & s2 & s1 & s4 & s3 & s6 & s5 & 0 & 0 & 0 & 0 & 0 & 0 & 0 & 0 & 0 & 0
   & 0 & 0 & 0 & 0 & 0 & 0 & 0 & 0 & 0 \cr 0 & 0 & 0 & 0 & 0 & 0 & 0 & 0 & 0
   & 0 & 0 & 0 & 0 & 0 & 0 & 0 & 0 & 0 & 0 & 0 & 0 & 0 & 0 & 0 & 0 & 0 & 0
   \cr 0 & s1 & 0 & 0 & 0 & 0 & 0 & 0 & 0 & 0 & 0 & 0 & 0 & 0 & 0 & 0 & 0 & 0
   & 0 & 0 & 0 & 0 & 0 & 0 & 0 & 0 & 0 \cr 0 & s2 & 0 & 0 & 0 & 0 & 0 & 0 & 0
   & 0 & 0 & 0 & 0 & 0 & 0 & 0 & 0 & 0 & 0 & 0 & 0 & 0 & 0 & 0 & 0 & 0 & 0
   \cr 0 & s3 & 0 & 0 & 0 & 0 & 0 & 0 & 0 & 0 & 0 & 0 & 0 & 0 & 0 & 0 & 0 & 0
   & 0 & 0 & 0 & 0 & 0 & 0 & 0 & 0 & 0 \cr 0 & s4 & 0 & 0 & 0 & 0 & 0 & 0 & 0
   & 0 & 0 & 0 & 0 & 0 & 0 & 0 & 0 & 0 & 0 & 0 & 0 & 0 & 0 & 0 & 0 & 0 & 0
   \cr 0 & s5 & 0 & 0 & 0 & 0 & 0 & 0 & 0 & 0 & 0 & 0 & 0 & 0 & 0 & 0 & 0 & 0
   & 0 & 0 & 0 & 0 & 0 & 0 & 0 & 0 & 0 \cr 0 & s6 & 0 & 0 & 0 & 0 & 0 & 0 & 0
   & 0 & 0 & 0 & 0 & 0 & 0 & 0 & 0 & 0 & 0 & 0 & 0 & 0 & 0 & 0 & 0 & 0 & 0
   \cr 0 & 0 & 0 & 0 & 0 & 0 & 0 & 0 & 0 & 0 & 0 & 0 & 0 & 0 & 0 & 0 & 0 & 0
   & 0 & 0 & 0 & 0 & 0 & 0 & 0 & 0 & 0 \cr 0 & 0 & 0 & 0 & 0 & 0 & 0 & 0 & 0
   & 0 & 0 & 0 & 0 & 0 & 0 & 0 & 0 & 0 & 0 & 0 & 0 & 0 & 0 & 0 & 0 & 0 & 0
   \cr 0 & 0 & 0 & 0 & 0 & 0 & 0 & 0 & 0 & 0 & 0 & 0 & 0 & 0 & 0 & 0 & 0 & 0
   & 0 & 0 & 0 & 0 & 0 & 0 & 0 & 0 & 0 \cr 0 & 0 & s13 & 0 & s14 & 0 & s15 & 0 &
  0 & 0 & 0 & 0 & 0 & 0 & 0 & 0 & 0 & s1 & s3 & s5 & s7 & 0 & 0 & 0 & 0 & 0 & 0
   \cr 0 & 0 & 0 & s9 & 0 & s10 & 0 & s11 & 0 & 0 & 0 & 0 & 0 & s2 & s4 & s6 & s8 & 0
   & 0 & 0 & 0 & 0 & 0 & 0 & 0 & 0 & 0 \cr 0 & s9 & 0 & 0 & 0 & 0 & 0 & 0 & 0
   & 0 & 0 & 0 & 0 & 0 & 0 & 0 & 0 & 0 & 0 & 0 & 0 & 0 & 0 & 0 & 0 & 0 & 0
   \cr 0 & s10 & 0 & 0 & 0 & 0 & 0 & 0 & 0 & 0 & 0 & 0 & 0 & 0 & 0 & 0 & 0 & 0
   & 0 & 0 & 0 & 0 & 0 & 0 & 0 & 0 & 0 \cr 0 & s11 & 0 & 0 & 0 & 0 & 0 & 0 & 0
   & 0 & 0 & 0 & 0 & 0 & 0 & 0 & 0 & 0 & 0 & 0 & 0 & 0 & 0 & 0 & 0 & 0 & 0
   \cr 0 & s12 & 0 & 0 & 0 & 0 & 0 & 0 & 0 & 0 & 0 & 0 & 0 & 0 & 0 & 0 & 0 & 0
   & 0 & 0 & 0 & 0 & 0 & 0 & 0 & 0 & 0 \cr 0 & s13 & 0 & 0 & 0 & 0 & 0 & 0 & 0
   & 0 & 0 & 0 & 0 & 0 & 0 & 0 & 0 & 0 & 0 & 0 & 0 & 0 & 0 & 0 & 0 & 0 & 0
   \cr 0 & s14 & 0 & 0 & 0 & 0 & 0 & 0 & 0 & 0 & 0 & 0 & 0 & 0 & 0 & 0 & 0 & 0
   & 0 & 0 & 0 & 0 & 0 & 0 & 0 & 0 & 0 \cr 0 & s15 & 0 & 0 & 0 & 0 & 0 & 0 & 0
   & 0 & 0 & 0 & 0 & 0 & 0 & 0 & 0 & 0 & 0 & 0 & 0 & 0 & 0 & 0 & 0 & 0 & 0
   \cr 0 & s16 & 0 & 0 & 0 & 0 & 0 & 0 & 0 & 0 & 0 & 0 & 0 & 0 & 0 & 0 & 0 & 0
   & 0 & 0 & 0 & 0 & 0 & 0 & 0 & 0 & 0 \cr 0 & 0 & 0 & s14 & 0 & s13 & s12 & 0 &
  0 & 0 & 0 & 0 & 0 & 0 & 0 & s7 & s5 & s4 & s2 & 0 & 0 & 0 & 0 & 0 & 0 & 0 & 0
   \cr 0 & 0 & 0 & s15 & s12 & 0 & 0 & s13 & 0 & 0 & 0 & 0 & 0 & 0 & s7 & 0 & s3 &
  6 & 0 & s2 & 0 & 0 & 0 & 0 & 0 & 0 & 0 \cr 0 & 0 & 0 & s16 & s11 & 0 & s10 & 0
   & 0 & 0 & 0 & 0 & 0 & 0 & s5 & s3 & 0 & s8 & 0 & 0 & s2 & 0 & 0 & 0 & 0 & 0 & 0
   \cr 0 & 0 & s12 & 0 & 0 & s15 & 0 & s14 & 0 & 0 & 0 & 0 & 0 & s7 & 0 & 0 & s1 &
  0 & s6 & s4 & 0 & 0 & 0 & 0 & 0 & 0 & 0 \cr 0 & 0 & s11 & 0 & 0 & s16 & s9 & 0 &
  0 & 0 & 0 & 0 & 0 & s5 & 0 & s1 & 0 & 0 & s8 & 0 & s4 & 0 & 0 & 0 & 0 & 0 & 0
   \cr 0 & 0 & s10 & 0 & s9 & 0 & 0 & s16 & 0 & 0 & 0 & 0 & 0 & s3 & s1 & 0 & 0 & 0
   & 0 & s8 & s6 & 0 & 0 & 0 & 0 & 0 & 0 \cr  }
\end{equation}}
The maximal number of gaugeable translational isometries is 12.

We list in the following, with the same notations as before, the analogous matrices in $D=6,7,8,9$,
which have dimensions 16, 10, 6 and 3 respectively.We number the rows
and columns according to eq.s\eqn{d5root},\eqn{d4root},\eqn{d3root}
and \eqn{e2root}. (In the last case, corresponding to D=9 there are
two additional vector fields besides the one corresponding to the
$E_2$ root.
\par
In each case the number of gaugeable translational isometries turns out to be 6,4,3,1 respectively.

$D=6$:
{\scriptsize
\begin{equation}
\label{rap4}
\pmatrix{ 0 & 0 & s2 & s1 & s4 & s3 & s6 & s5 & 0 & 0 & 0 & 0 & 0 & 0 & 0 & 0 \cr 0
   & 0 & 0 & 0 & 0 & 0 & 0 & 0 & 0 & 0 & 0 & 0 & 0 & 0 & 0 & 0 \cr 0 & s1 & 0
   & 0 & 0 & 0 & 0 & 0 & 0 & 0 & 0 & 0 & 0 & 0 & 0 & 0 \cr 0 & s2 & 0 & 0 & 0
   & 0 & 0 & 0 & 0 & 0 & 0 & 0 & 0 & 0 & 0 & 0 \cr 0 & s3 & 0 & 0 & 0 & 0 & 0
   & 0 & 0 & 0 & 0 & 0 & 0 & 0 & 0 & 0 \cr 0 & s4 & 0 & 0 & 0 & 0 & 0 & 0 & 0
   & 0 & 0 & 0 & 0 & 0 & 0 & 0 \cr 0 & s5 & 0 & 0 & 0 & 0 & 0 & 0 & 0 & 0 & 0
   & 0 & 0 & 0 & 0 & 0 \cr 0 & s6 & 0 & 0 & 0 & 0 & 0 & 0 & 0 & 0 & 0 & 0 & 0
   & 0 & 0 & 0 \cr 0 & 0 & 0 & s7 & 0 & s8 & 0 & s9 & 0 & s2 & s4 & s6 & 0 & 0 & 0
   & 0 \cr 0 & s7 & 0 & 0 & 0 & 0 & 0 & 0 & 0 & 0 & 0 & 0 & 0 & 0 & 0 & 0 \cr 0
   & s8 & 0 & 0 & 0 & 0 & 0 & 0 & 0 & 0 & 0 & 0 & 0 & 0 & 0 & 0 \cr 0 & s9 & 0
   & 0 & 0 & 0 & 0 & 0 & 0 & 0 & 0 & 0 & 0 & 0 & 0 & 0 \cr 0 & s10 & 0 & 0 & 0
   & 0 & 0 & 0 & 0 & 0 & 0 & 0 & 0 & 0 & 0 & 0 \cr 0 & 0 & 0 & s10 & s9 & 0 & s8
   & 0 & 0 & 0 & s5 & s3 & s2 & 0 & 0 & 0 \cr 0 & 0 & s9 & 0 & 0 & s10 & s7 & 0 & 0
   & s5 & 0 & s1 & s4 & 0 & 0 & 0 \cr 0 & 0 & s8 & 0 & s7 & 0 & 0 & s10 & 0 & s3 & s1
   & 0 & s6 & 0 & 0 & 0 \cr  }
 \end{equation}}
$D=7$:
\begin{equation}
\label{rap3}
\pmatrix{ 0 & 0 & s2 & s1 & s4 & s3 & 0 & 0 & 0 & 0 \cr 0 & 0 & 0 & 0 & 0 & 0 & 0
   & 0 & 0 & 0 \cr 0 & s1 & 0 & 0 & 0 & 0 & 0 & 0 & 0 & 0 \cr 0 & s2 & 0 & 0 & 0
   & 0 & 0 & 0 & 0 & 0 \cr 0 & s3 & 0 & 0 & 0 & 0 & 0 & 0 & 0 & 0 \cr 0 & s4 & 0
   & 0 & 0 & 0 & 0 & 0 & 0 & 0 \cr 0 & 0 & 0 & s5 & 0 & s6 & 0 & s2 & s4 & 0 \cr 0
   & s5 & 0 & 0 & 0 & 0 & 0 & 0 & 0 & 0 \cr 0 & s6 & 0 & 0 & 0 & 0 & 0 & 0 & 0
   & 0 \cr 0 & 0 & s6 & 0 & s5 & 0 & 0 & s3 & s1 & 0 \cr  }
\end{equation}

$D=8$:
\begin{equation}
\label{rap2}
\pmatrix{ 0 & 0 & s2 & s1 & 0 & 0 \cr 0 & 0 & 0 & 0 & 0 & 0 \cr 0 & s1 & 0 & 0 & 0
   & 0 \cr 0 & s2 & 0 & 0 & 0 & 0 \cr 0 & 0 & 0 & s3 & 0 & s2 \cr 0 & s3 & 0 & 0
   & 0 & 0 \cr  }
\end{equation}

$D=9$:
\begin{equation}
\label{rap1}
\pmatrix{ 0 & 0 & s1 \cr 0 & 0 & 0 \cr 0 & 0 & 0 \cr   }
\end{equation}


 \end{document}